\documentclass{pazha2}
\usepackage{placeins}
\usepackage{ulem}
\usepackage{graphicx}
\usepackage{multirow}
\usepackage{color}
\definecolor{darkblue}{rgb}{0,0,0.9}

\def\*{$^{*}$}
\def\aa{$^{\mbox{\small a}}$}
\def\bb{$^{\mbox{\small b}}$}
\def\cc{$^{\mbox{\small c}}$}
\def\dd{$^{\mbox{\small d}}$}
\def\ee{$^{\mbox{\small e}}$}
\def\ff{$^{\mbox{\small f}}$}

\def\з{$^{\mbox{\small з}}$}
\def\и{$^{\mbox{\small и}}$}
\def\к{$^{\mbox{\small к}}$}

\begin{document}
\journalinfo{to be}{2024}{50}{3}{159}{175}{200}[185]
\sloppypar

\title{\bf INCREASE IN THE BRIGHTNESS OF THE COSMIC RADIO
  BACKGROUND TOWARD GALAXY CLUSTERS}
\year=2024
\author{S. A.~Grebenev\email{grebenev@iki.rssi.ru}\address{1} and
R. A.~Sunyaev\address{1,2}
\addresstext{1}{Space Research
    Institute, Russian Academy of Sciences, Profsoyuznaya ul. 
    84/32, Moscow, 117997 Russia}
\addresstext{2}{Max-Planck-Institut f\"{u}r Astrophysik,
  Karl-Schwarzschild-Str.\,1, Postfach\,1317, D-85741 Garching,\,Germany}
}

\shortauthor{GREBENEV, SUNYAEV}
\shorttitle{INCREASE IN THE BRIGHTNESS OF THE COSMIC RADIO BACKGROUND} 

\submitted{December 15, 2022}
\revised{April 18, 2023}
\accepted{June 2, 2023}

\begin{abstract}
  \noindent\footnotesize We explore the possibility of detecting
  the excess of the cosmic radio background toward galaxy
  clusters due to its Compton scattering by electrons of the hot
  intergalactic gas. When mapping the background fluctuations at
  frequencies below $\la 800$ MHz, this effect gives rise to a
  radio source at the location of the cluster. At higher
  frequencies, where the microwave (relic) radiation dominates
  in the cosmic background, a ``negative'' source (a ``shadow''
  on the map of background fluctuations) is observed at the
  location of the cluster due to the transfer of some of the
  relic photons upward along the frequency axis upon their
  scattering (into the range $\nu\ga 217$ GHz; Sunyaev and
  Zeldovich 1970, 1972).  We have computed the spectra of the
  expected radio background distortions for various parameters
  of clusters and show that in many cases in the wide frequency
  range $30\ \mbox{\rm MHz}\la \nu\la 3\ \mbox{\rm GHz}$ the
  measurement of the distortions will be hindered by the
  intrinsic thermal (bremsstrahlung) radiation from the
  intergalactic gas and the scattered radio emission from
  cluster galaxies associated with their past activity,
  including the synchrotron radiation from ejected relativistic
  electrons. Below $\sim20$ MHz the scattering effect always
  dominates over the thermal gas radiation due to the general
  increase in the intensity of the cosmic radio background, but
  highly accurate measurements at such frequencies become
  difficult. Below $\sim5$ MHz the effect is suppressed by the
  induced scattering. We have found the frequency ranges that
  are optimal for searching for and measuring the Compton radio
  background excess. We show that hot ($kT_{\rm e} \ga
  8\ \mbox{\rm keV}$) clusters at high ($z \ga 0.5$) redshifts
  are most promising for its observation. Because of the strong
  concentration of the bremsstrahlung to the cluster center, the
  peripheral observations of the Compton excess must be more
  preferable than the central ones.  Moreover, owing to the
  thermal radiation of the gas and its concentration to the
  center, the above-noted transition from the ``negative''
  source on the map of background fluctuations to the
  ``positive'' one when moving downward along the frequency axis
  must occur not gradually but through the stage of a ``hybrid
  source'' --- the appearance of a bright spot surrounded by a
  dark ring. This shape of the source in projection is explained
  by its unusual three-dimensional shape in the form of a narrow
  radio bremsstrahlung peak rising from the center of a wide
  deep hole associated with the Compton scattering of the cosmic
  microwave background. The scattered radiation from a
  central cluster galaxy active in the past can amplify the effect. An
  analogous ``hybrid source'' also appears on the map of
  background fluctuations near a frequency of $217.5$ GHz ---
  when passing from the deficit of the cosmic microwave
  background to its excess (due to the scattered photons). The
  unusual shape of the source is again associated with the
  thermal gas radiation. Simultaneous measurements of the radio
  bremsstrahlung flux from the gas and the amplitude of the
  distortions due to the radio and cosmic microwave background
  scattering will allow the most important cluster parameters to
  be determined.\\

\noindent
{\bf DOI:} 10.1134/S1063773724700063

\end{abstract}
\section{INTRODUCTION}
\noindent
The effect of a decrease in the brightness of the microwave
background toward galaxy clusters (Sunyaev and Zeldovich 1970,
1972, 1980, 1981; Zeldovich and Sunyaev 1982) is widely used to
investigate the properties of clusters and other objects in the
early Universe and their evolution. The decrease in brightness
is associated with the deficit of cosmic microwave background
(CMB) photons relative to the Planck spectrum at frequencies
below $\nu_0\simeq 217$ GHz due to their upward shift along the
frequency axis upon Compton scattering by electrons of the hot
($kT_{\rm e}\sim 3-15$ keV) intergalactic cluster gas. Here,
$h\nu_0 \simeq 3.83\,kT_{\rm m}$, where $T_{\rm m}=2.7255\ K$ is
the current CMB temperature. A ``shadow'' (a ``negative''
source) appears on the microwave background map toward a
cluster. An excess of photons is formed at frequencies $\nu\ga
\nu_0$, and a ``positive'' source flares up on the background
map. The action of the effect is determined by the optical depth
of the cluster gas for scattering by electrons $\tau_{\rm
  T}=\sigma_{\rm T}\int N_{\rm e}(l)\,{\rm d}\,l$, i.e.,
proportionally to the gas density along the line of sight and
not to the square of the density (like the brightness of the
thermal gas radiation). Here, $\sigma_{\rm T}$ is the Thomson
scattering cross section and $N_{\rm e}(l)$ is the electron
density. The amplitude of the effect (the drop in the spectral
flux density) does not decrease with distance to the cluster
(its redshift $z$); the shape of the background distortion
spectrum does not depend on $z$ either. Owing to these
properties, the effect is widely used to determine the
parameters of known clusters and to search for new clusters.

The effect has been successfully observed with the specially
built South Pole Telescope (SPT, Carlstrom et al. 2002;
Williamson et al. 2011; Bleem et al., 2015, 2020) and Atacama
Cosmology Telescope (ACT, Hasselfield et al. 2013; Hilton et
al. 2021) as well as a number of other telescopes (Birkinshaw
1999); the Planck satellite (Ade et al. 2014, 2015, 2016) has
contributed enormously to the investigation of the
effect. Several more specialized instruments and telescopes must
join the extensive studies of the effect in the near future (for
a review, see Mroczkowski et al. 2019).

The samples of galaxy clusters detected through the effect turn
out to be much more representative at high ($z\ga 0.5$)
redshifts than the samples of clusters found from X-ray
data. Therefore, the dependences of the number of clusters on
their mass and $z$ derived from such samples are effectively
used to obtain constraints on the parameters of the cosmological
models of the Universe (see, e.g., de Haan et al. 2016).
\begin{figure}[!h]
\hspace{-5mm}\includegraphics[width=1.09\linewidth]{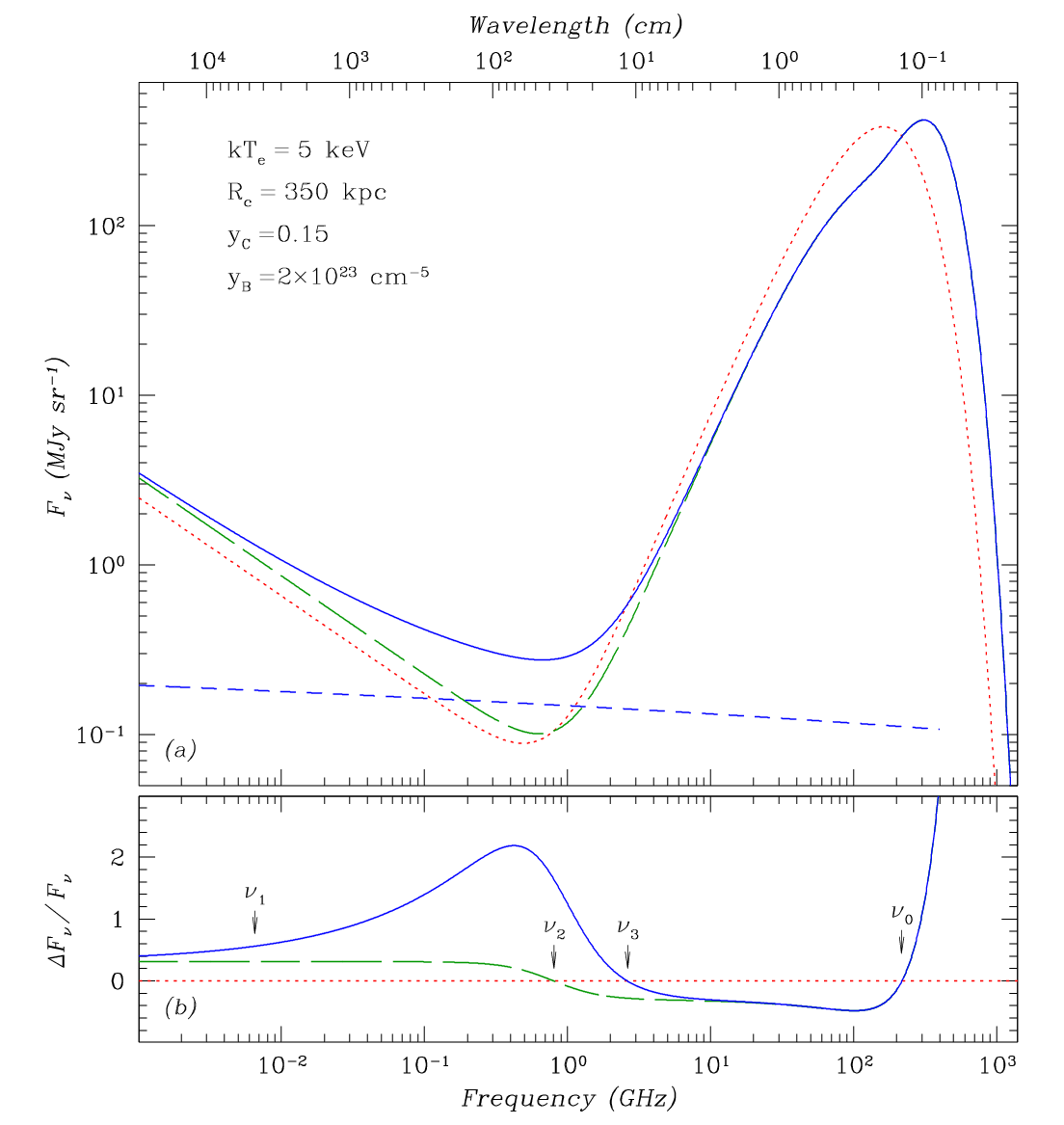}

\caption{\rm {\sl (a)\/} The radio and microwave background
  spectrum (red dotted line) and the corresponding distorted
  spectrum due to the scattering by electrons of the hot gas in
  a galaxy cluster (long green dashes) and the contribution of
  the bremsstrahlung from this gas (blue solid line, the
  spectrum of the bremsstrahlung itself is indicated by the
  dashed straight line). The demonstration computation for a
  hypothetical cluster with a uniform density distribution, a
  radius $R_{\rm c}=350$ kpc, a temperature $kT_{\rm e}=5$ keV,
  and Compton, $y_{\rm C}=0.15$, and bremsstrahlung, $y_{\rm
    B}=2\times 10^{23}\ \mbox{cm}^{-5}$, parameters that
  actually define the distortion amplitudes (for real clusters
  $y_{\rm C}$ and $y_{\rm B}$ have values that are lower by
  three orders of magnitude). {\sl (b)\/} The relative
  distortions in the background spectrum toward this cluster
  (the blue solid curve takes into account the gas
  bremsstrahlung). The frequencies $\nu_2\simeq 802$ MHz (of the
  equality of the Compton radio and microwave background
  distortions in absolute value), $\nu_1$ and $\nu_3$ (of the
  equality of the bremsstrahlung flux and the Compton excess in
  the radio background spectrum or the Compton dip in the CMB
  spectrum), and $\nu_0=217$ GHz (of the transition from the
  deficit of photons to their excess in this spectrum) are
  indicated.\label{fig:bkg}}
\end{figure}

The possibility of observing such an effect in other wavelength
ranges is widely discussed. For example, Grebenev and Sunyaev
(2019) computed the distortions that appear in the spectrum of
the cosmic X-ray and soft gamma-ray background upon its Compton
scattering and photoabsorption in the hot gas of galaxy
clusters. Sabyr et al. (2022) investigated the distortions that
arise upon inverse Compton scattering in the cosmic infrared
background spectrum.  Cooray (2006) considered the distortions
that arise upon scattering by electrons in the 21 cm background
profile. Quite recently, Holder and Chluba (2021) and Lee et
al. (2022) considered analogous distortions already in the radio
background continuum itself.

The cosmic radio background dominates over the CMB at
frequencies below $\sim1$ GHz. It was discovered in the ARCADE 2
balloon experiment (Fixsen et al. 2011) during highly accurate
measurements and was initially discussed as the mysterious
``ARCADE excess''. The ARCADE 2 measurements were supplemented
by the radar observations at low frequencies of 22 and 45 MHz
(Maeda et al. 1999; Roger et al. 1999), the LLFSS sky survey at
frequencies 40--80 MHz (Dowell and Taylor 2018), and the surveys
at 408 MHz (Remazeilles et al. 2015) and 1.4 GHz (Reich et
al. 2001).The radio background was shown to have a power-law
synchrotron spectrum with a spectral index $\alpha\simeq
0.58\pm0.05$
in a wide wavelength range. The nature of the background is
still unknown; it is possible to associate no more than 25\% of
the background radiation with radio galaxies, active galactic
nuclei, and other faint compact sources (Seiffert et al. 2011;
Condon et al. 2012; Hardcastle et al. 2021). Other reasons for
its existence being discussed do not appear convincing either
(see Singal et al. 2023). In any case, as it should be for the
background, this radio emission is characterized by a high
degree of isotropy and homogeneity.

The estimates by Holder and Chluba (2021) showed that the
Compton scattering of the radio background by electrons of the
hot cluster gas increases its brightness at all frequencies
(the change in the brightness temperature $\Delta T$ reaches
$\sim\,1\,\mbox{\rm mK}$).  Near $\nu_2\simeq 802$ MHz this
increase completely compensates for the above-mentioned
decrease in the brightness of the microwave background, i.e., at
frequencies $\nu\la\nu_2$ the ``source'' reappears on the
background map toward the cluster instead of the ``shadow''.
    
In Fig.~\ref{fig:bkg}a the background spectrum distorted in the
cluster gas that was computed according to Holder and Chluba
(2021) is indicated by the long green dashes. The undistorted
spectrum (the sum of the cosmic radio and microwave backgrounds)
is indicated by the red dotted line. The cluster Compton
parameter that defines the amplitude of the spectral distortions
is greatly overestimated here, $y_{\rm C}=\sigma_{\rm T}\int
N_{\rm e}(l)\,(kT_{\rm e}/m_{\rm e}c^2)\,{d}\,l=0.15,$ but, on
the whole, the figure conveys correctly the action of the Compton
scattering on the background spectrum.  In typical galaxy
clusters the parameter $y_{\rm C}$ has a much lower value. For
example, in a cluster with an electron temperature $kT_{\rm
  e}=5$ keV and an optical depth of the gas toward the cloud
center $\tau_{\rm T}=0.01$ it is $y_{\rm C}=1\times10^{-4}.$

In this paper we show that the effect predicted by Holder and
Chluba (2021) and more rigorously calculated by Lee et
al. (2021) is unobservable in most clusters. In the decimeter,
meter, and decameter wavelength ranges (frequencies $\nu\la\,5$
GHz) the intrinsic bremsstrahlung of the hot intergalactic gas
in such clusters exceeds noticeably in flux and completely
suppresses the Compton radio and microwave background
distortions. This is illustrated by Fig.\,\ref{fig:bkg}a, in
which the background spectrum (long green dashes) including the
bremsstrahlung (short blue dashes) is indicated by the blue
solid line. The intrinsic gas radiation is seen to dominate in
the overall spectrum of the radio emission detected toward the
cluster, and only at frequencies below $\nu_1\simeq 10$ MHz does
its flux become equal to the Compton radio background
distortions (due to the general increase in the background
brightness).

This can be seen even better in Fig.\,\ref{fig:bkg}b that
presents the relative background distortions toward this cluster
(with and without the contribution from the intergalactic gas
bremsstrahlung). Note that, for demonstration purposes, the gas
emission measure toward the center of this hypothetical cluster
was greatly overestimated, $y_{\rm B}=\Sigma_z \int Z^2 N_{\rm
  z}(l) N_{\rm e}(l)\,{d}\,l=2\times 10^{23}\ \mbox{cm}^{-5}$
(the summation is over the plasma ions, $Z$ is the nuclear
charge of the ion). In reality, it is usually lower by two or
three orders of magnitude ($y_{\rm B}=1.2\times
10^{20}\ \mbox{cm}^{-5}$ for the mentioned cluster with
$\tau_{\rm T}=0.01$). Below in the paper we will calculate the
exact values of the above-mentioned frequency $\nu_1$ and the
frequency $\nu_3$ at which the gas bremsstrahlung flux completely
compensates for the Compton decrease in the CMB brightness for
realistic parameters of clusters.

Given the smallness of the parameters $y_{\rm C}$ and $y_{\rm
  B}$ (a factor of 1000 lower than the adopted values) for
typical galaxy clusters, Fig.\,\ref{fig:bkg}b implies that the
distortions in the radio background (and centimeter CMB)
spectrum considered in the paper are small in absolute value
(being fractions of a percent of the background level) and are
at the sensitivity limit of modern telescopes. Nevertheless, the
rapid development of radio astronomy, as new radio telescopes
and radio interferometers are put into operation, such as GMRT
(Venturi et al. 2008), LOFAR (van Haarlem et al. 2013), MeerKAT
(Jonas et al. 2016), ALMA/ACA (Di Mascolo 2020), ASKAP (Hotan et
al. 2021), SKA (Bacon et al. 2020), CHIME (Amiri et al. 2021),
and Tianlei DPA (Wu et al. 2021), promises a noticeable increase
in the accuracy and sensitivity of radio measurements and the
ability to map the sky at various frequencies with a high
angular resolution in the near future. This guarantees the
possibility of measuring the subtle effects being discussed in
the paper.

\section*{THE BACKGROUND SPECTRUM AND ITS DISTORTIONS}
\noindent
Recall the main physical processes that lead to distortions in
the spectrum of the background radiation as it interacts with
the hot intergalactic gas of a cluster.

\subsubsection*{\uwave{Compton scattering.}\/}~We will consider the
interaction of the background radiation with high-temperature
electrons in the gas of galaxy clusters by solving the
Kompaneets (1957) equation that describes the photon frequency
redistribution in the diffusion approximation. The distortions
in the CMB spectrum arising in clusters were previously
estimated in this way (Sunyaev and Zeldovich 1980; Zeldovich and
Sunyaev 1982). The applicability of this equation in the case of
an optically thin gas typical for clusters was checked and
confirmed by Sunyaev (1980).

The Kompaneets equation is
\begin{equation}\label{eq:comp}
  \frac{\partial F_{\nu}}{\partial\tau_{\rm T}}=\frac{h\nu}{m_{\rm e}c^2}
  \frac{\partial}{\partial\nu}\left[
    \nu^4\left(\frac{F_{\nu}}{\nu^3}+
    \frac{c^2}{2h}\frac{F_{\nu}^2}{\nu^6}+
    \frac{kT_{\rm e}}{h}\frac{\partial}{\partial\nu}
    \frac{F_{\nu}}{\nu^3}\right)\right]\hspace{-1mm}
  \end{equation}   
Here $F_{\nu}$ is the spectral intensity of the background
radiation. The nonlinear term $\sim F_{\rm \nu}^2$ on the
right-hand side of the equation takes into account the induced
scattering, the first term $\sim F_{\rm \nu}$ is responsible for
the recoil effect, it is a factor of $\sim kT_{\rm e}/h\nu \sim
10^{10}$ smaller than the last (Doppler) term.

\subsubsection*{\uwave{Distortions in the CMB spectrum.}\/}~Neglecting
the first two terms, let us substitute the Planck spectrum of the
microwave background, $$B_{\nu}(\nu)=\frac{2h\nu^3}{c^2}\,\left[\exp\left(\frac{h\nu}{kT_{\rm
    m}}\right)-1\right]^{-1},$$ into Eq.~(\ref{eq:comp}). We obtain the
well-known spectrum of the CMB distortions toward the cluster
center
\begin{equation}\label{eq:szcmb}
  \frac{\Delta B_{\nu}}{B_{\nu}}= y_{\rm C}\,\frac{xe^x}{e^x-1}\,\left[x\, \left(\frac{e^x+1}{e^x-1}\right)-4\right].
\end{equation}
Here, $x=h\nu/kT_{\rm m}$ and $y_{\rm C}=\tau_{\rm T}\,(kT_{\rm
  e}/m_{\rm e} c^2)$ is the previously introduced Compton
parameter that defines the distortion amplitude. In the limit
$x\ll 1$ ($\nu\ll 57$ GHz) the relative distortions of the
Planck spectrum are negative and do not depend on the frequency,
$$\Delta B_{\nu}/B_{\nu}\simeq -2 y_{\rm C}.$$

\subsubsection*{\uwave{Distortions in the radio background spectrum.}\/}~According to Fixsen et al. (2011) and Dowell and
Taylor (2018), the intensity of the radio background at our
($z=0$) epoch is a power-law function of frequency, in terms of
the brightness temperature
\begin{equation}\label{eq:tcrb}
  T_{R}(\nu)=T_*\, (\nu/\nu_*)^{-2.58\pm0.05}\, \mbox{K}, 
\end{equation}
where $\nu_*=310$ MHz, $T_*=(30.4\pm 2.1)\,
\mbox{K}$. Substituting the intensity of the radiation in the
form $F_{R}(\nu)=F_*\nu^{-\alpha},$ where
$\alpha=0.58\pm0.05,$ into the right-hand side of
Eq.~(\ref{eq:comp}), we find the spectrum of its relative
distortions
\begin{equation}\label{eq:szcrb}
\Delta\,F_{R}/F_{R}= y_{\rm C}\ \alpha (3+\alpha)\simeq 2.08\,y_{\rm C}.
\end{equation}
Thus, the relative amplitude of the effect for the radio
background is again proportional to the Compton pa- rameter
$y_{\rm C}$, but, at the same time, it is always positive and
does not depend on the frequency $\nu$ (Holder and Chluba 2021).

The distortions of the microwave, $\Delta B_{\nu}$, and radio,
$\Delta F_{\rm R}$, background fluxes become equal in absolute
value (compensate each other) at a frequency
$$\nu_2=\nu_*\ [0.5 (3+\alpha)\, \alpha T_*/T_m ]^{1/(\alpha+2)}\simeq
802\pm38\ \mbox{MHz},
$$ which does not depend on the cluster gas parameters (Holder
and Chluba 2021).
\subsubsection*{\uwave{Distortions due to the induced scattering.}\/}~The intensity of the radio background at frequencies
$\nu\la2.5$ GHz corresponds to an occupation number $n_{\rm
  \nu}(\nu) = F_{R}(\nu)\,c^2/2h\nu^3\simeq
2045\ (\nu_*/\nu)^{(3+\alpha)}\ga 1$ and, hence, the term in
Eq.~(\ref{eq:comp}) responsible for the induced Compton
scattering cannot be neglected a priori. Retaining this term and
substituting again the intensity of the radiation in the form
$F_{R}(\nu) = F_* \nu^{-\alpha}$ into the right-hand side of
the equation, we obtain
\begin{equation}\label{eq:szcri}
\Delta\,F_{R}/F_{R}=y_{\rm C}\, \alpha
(3+\alpha)-\tau_{\rm T} (1+\alpha)/m_{\rm e} F_{*} \nu^{-(2+\alpha)}
\end{equation}
The contribution of the induced Compton scattering has a
negative sign. This is natural, since it leads to the downward
shift of low-frequency photons along the frequency axis (Sunyaev
1970). This contribution becomes equal in absolute value to the
Doppler term at $\nu_{\rm 4}\simeq 1.3\ (kT_{\rm e}/5\, 
\mbox{\rm keV})^{-0.39}$ MHz, at lower frequencies the
brightness of the radio background turns out to be reduced
again.

\subsubsection*{\uwave{Bremsstrahlung of the intergalactic gas.}\/}~The
brems\-strah\-lung surface brightness of an isothermal hot gas in a
galaxy cluster toward its center is (Lang 1974)
\begin{equation}\label{eq:brem}
 F_{B}(\nu) = A\, \frac{\int \Sigma_z (Z^2\,N_{\rm Z})
   N_{\rm e}\,dl}{T_{\rm e}^{1/2}}\, g(\nu,T_{\rm
   e})\exp{\left(-\frac{h\nu}{kT_{\rm e}}\right)}
\end{equation}
where the constant $A=$
$$
=\frac{8}{3}\left(\frac{2\pi}{3}\right)^{1/2}\!\!\frac{e^6k^{-1/2}}{(m_{\rm e}c^2)^{3/2}}\simeq
5.4\times
 10^{-39}\ \frac{\mbox{erg cm}^{3}\,\mbox{K}^{1/2}}{\mbox{s Hz sr}},
$$
and the Gaunt factor $$g(\nu, T_{\rm
  e})=\frac{\sqrt{3}}{\pi}\ln{\,\Lambda},\ \mbox{where}\ \Lambda=\frac{4kT_{\rm
    e}}{\gamma\,h\nu}\simeq4.7\times 10^{10} 
\left(\frac{T_{\rm e}}{\nu}\right)
$$ ($\gamma\simeq 1.781$, $T_{\rm e}$ is in K, $\nu$ is in Hz).
Using the previously introduced bremsstrahlung parameter of the
cluster gas (the gas emission measure along the line of sight)
$y_{\rm B}$, Eq.~(\ref{eq:brem}) can be represented as
\begin{equation}\label{eq:brem2}
   F_{B}(\nu) =y_{\rm B}\ A T_{\rm e}^{-1/2}\,
 g(\nu,T_{\rm e})\exp{(-h\nu/kT_{\rm e}).}  
\end{equation}
By comparing Eqs. (\ref{eq:brem2}) and (\ref{eq:szcrb}), we can
find the frequency $\nu_1,$ at which the radio background
distortions $\Delta F_{R}$ and the contribution of the
bremsstrahlung $F_{B}$ become equal. Similarly, by comparing
Eqs.~(\ref{eq:brem2}) and (\ref{eq:szcmb}), we can find the
frequency $\nu_3$ at which the microwave background distortions
$\Delta B_{\nu}$ in the limit $\nu\ll 57$ GHz become equal in
absolute value to the contribution of the bremsstrahlung $F_{B}$
(they compensate each other). In contrast to the frequency
$\nu_2$, these frequencies are not universal and depend in a
certain way on the intergalactic gas temperature and density.

\subsubsection*{\uwave{Bremsstrahlung absorption by the intergalactic gas.}\/}~The bremsstrahlung processes also
lead to the absorption of the radio background at low
frequencies. The optical depth for this process toward the
cluster center is (Lang 1974)
\begin{equation}\label{eq:brabs}
 \tau_{\rm B}(\nu)=\frac{F_{B}(\nu)}{2kT_{\rm e}} \frac{c^2}{\nu^2}
 =y_{\rm B} \frac{A c^2}{2k\nu^2}\ g(\nu, T_{\rm e})\, T_{\rm e}^{-3/2},  
\end{equation}
where $F_{B}(\nu)$ is substituted from
Eq.~(\ref{eq:brem2}). Accordingly, the distortion in the radio
background spectrum in addition to the Compton distortions
described by Eqs.~(\ref{eq:szcrb}) and (\ref{eq:szcri}) that is
related to the bremsstrahlung absorption in the intergalactic
gas is
\begin{equation}\label{eq:brabs2}
 \Delta F_{R}/F_{R}(\nu) = \exp[-\tau_{\rm
     B}(\nu)]\simeq 1-\tau_{\rm B}(\nu).
\end{equation}
The bremsstrahlung itself is also absorbed, but to a lesser
degree, since its intensity is gained already inside the cluster
(along the line of sight).

\section*{A MODEL CLUSTER}
\noindent
For clarity and simplicity, we will initially assume the cluster
gas to be distributed uniformly and to have the same
temperature.

\subsubsection*{\uwave{A cluster with a uniform density distribution.}\/}~Consider a spherically symmetric cloud of hot gas in a
cluster with an electron density $N_{\rm e}$ and temperature
$T_{\rm e}$ that has a radius $R_{\rm c}$. The optical depth of
such a cluster for Thomson scattering along the line of sight
passing through its center is $\tau_{\rm T}=2 \sigma_{\rm T}
N_{\rm e} R_{\rm c} = 2\tau_{\rm c}$. The electron density in
the gas cloud of a typical cluster with an optical depth
$\tau_{\rm T}=0.01$ and a radius $R_{\rm c}=350$ kpc is then
$N_{\rm e}\simeq 7.0\times10^{-3}\ \mbox{cm}^{-3}$. The gas mass
in the cloud is $M_{\rm g}\simeq 3.5\times10^{13}\ (\tau_{\rm
  T}\,/ 0.01)(r_{\rm c}/350\ \mbox{\rm kpc})^2\ M_{\odot}.$ The
total mass $M_{500}$ of the corresponding real (with dark
matter) cluster must be at least an order of magnitude
greater. This is a moderate-mass cluster like the Coma cluster.

Above and below we assume that hydrogen, helium, and oxygen in
the cluster gas have normal cosmic abundances by mass, $X\simeq
0.74,$ $Y\simeq 0.24$, and $O\simeq 0.01$ (Cameron 1982),
respectively, $N_{\rm e}\simeq(X+0.5\,Y+0.5\,O)\,\rho/m_{\rm
  p}\simeq 0.87\,\rho/m_{\rm p}$, the factor appearing in the
formula for calculating the bremsstrahlung parameter of the
intergalactic gas is $\Sigma\ Z^2 N_{\rm
  Z}\simeq(X+Y+4\,O)\,\rho/m_{\rm p}\simeq 1.02\,\rho/m_{\rm
  p}$, and this parameter itself is $y_{\rm B}=2 \Sigma Z^2
N_{\rm z} N_{\rm e} R_{\rm c}\simeq 2.36\ N_{\rm e}^2 R_{\rm
  c}$. At temperatures typical for the gas in galaxy clusters,
these elements are fully ionized. We will not take into account
the heavier elements.

We assume the background radiation to be incident on the cloud
isotropically. Accordingly, while computing the spectrum of the
radiation emerging from the cloud, we average it over the
angles. We assume the radio background, irrespective of its
nature (cosmological or associated with unresolved radio
galaxies), to be completely formed at high redshifts ($z>z_*$ of
the cluster). If this is not the case and some fraction of the
radio background is formed at $z<z_*$, then the amplitude of the
Compton distortions in its spectrum should be reduced
accordingly (by this fraction).
\begin{figure}[t]
\hspace{-5mm}\includegraphics[width=1.1\linewidth]{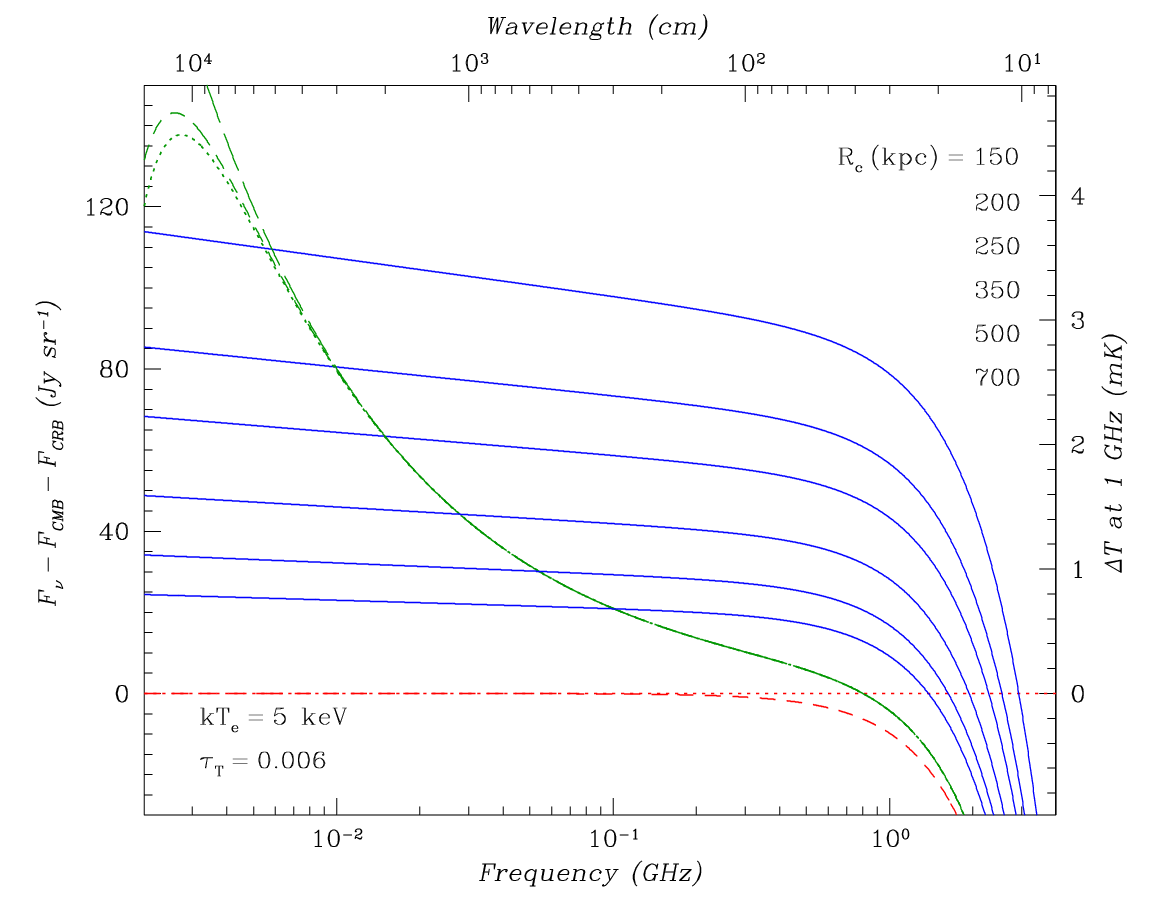}

\caption{\rm Comparison of the distortions in the radio
  background due to its scattering by electrons of the hot
  cluster gas (long green dashes) and due to the contribution
  from the bremsstrahlung of this gas (the blue solid lines
  corresponding to different cluster radii $R_{\rm c}$). In both
  cases, we took into account the CMB distortions due to the
  scattering by electrons (they are indicated separately by the
  short red dashes). The short green dashes indicate the
  decrease in the distortions due to the induced scattering, the
  green dotted line indicates their decrease also due to the
  bremsstrahlung absorption of radio emission in the gas. The
  computation for a nearby ($z\ll 1$) cluster with a uniform
  density distribution, a temperature $kT_{\rm e}=5$ keV, and a
  Thomson optical depth along the line of sight toward its
  center $\tau_{\rm T} = 6\times 10^{-3}$.
  \label{fig:dev1} } 
\end{figure}
\begin{figure}[t]
\hspace{-1mm}\includegraphics[width=1.10\linewidth]{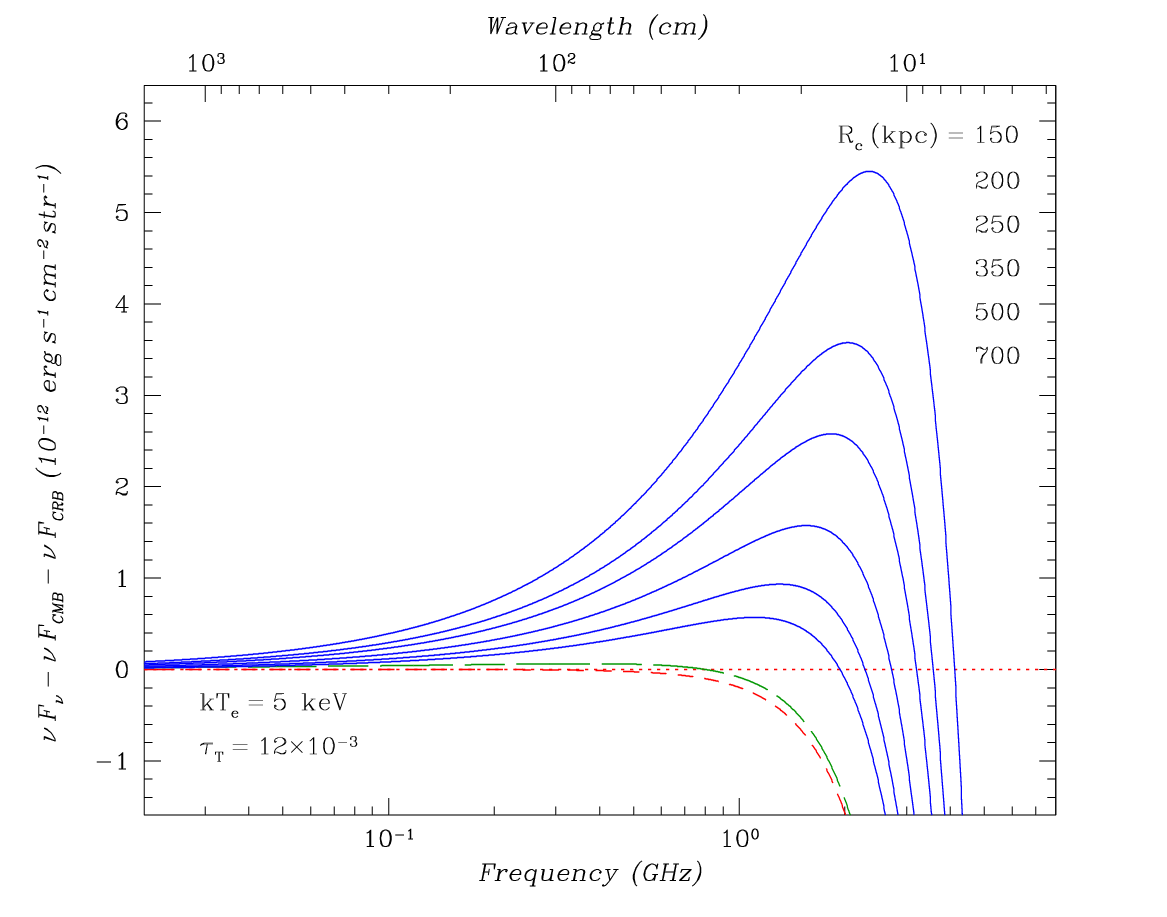}

\caption{\rm Same as Fig.\,\ref{fig:dev1}, but for the
  distortions in the spectrum $\nu\,F_{\nu}(\nu),$ which allows
  the contribution of the intergalactic gas bremsstrahlung to be
  investigated better in the centimeter-decimeter wavelength
  range, where the contribution of the Compton distortions in
  the CMB spectrum is great. For a gas temperature $kT_{\rm
    e}=5$ keV the bremsstrahlung is seen to compensate for the
  Compton drop in the flux up to $\lambda\sim 6$ cm ($\nu\sim 5$
  GHz). The optical depth of the gas is $\tau_{\rm T}=1.2\times
  10^{-2}$ (twice as large as that in 
  Fig.\,\ref{fig:dev1}).\label{fig:devnu}}
\end{figure}

\subsubsection*{\uwave{Computation of the background
    distortions.}\/}~The blue solid lines in
Fig.\,\ref{fig:dev1} indicate the contribution of the
intergalactic gas bremsstrahlung to the spectrum of the radio
and microwave background distortions that should be measured
toward the galaxy cluster (the measurements are assumed to be
carried out toward the cluster center). The cluster is assumed
to be a nearby one, located at a redshift $z\ll 1$. Different
lines correspond to different radii $R_{\rm c}$ of the gas cloud
(and, accordingly, different electron densities $N_{\rm e}$ in
it), the Thomson optical depth of the cloud along the line of
sight toward the cluster center is $\tau_{\rm T}=0.006$, and the
electron temperature is $kT_{\rm e}=5$ keV. The curves also take
into account the CMB distortion due to the scattering by
electrons of the hot cluster gas (the drop in flux at high
frequencies is associated precisely with it, it is indicated
separately by the short red dashes). The radio and microwave
background distortion due to the inverse Compton scattering by
hot-gas electrons is indicated by the long green dashes. The
same distortion including the induced scattering is indicated by
the short green dashes. The induced scattering shifts the
photons downward along the frequency axis, reducing the
amplitude of the Compton background distortions at very low
frequencies $\nu\la 10$ MHz. Yet another effect reducing the
distortion amplitude is the bremsstrahlung absorption of radio
emission in the hot cluster gas. It is indicated by the green
dotted line in addition to the contribution of the induced
scattering. In what follows, both these effects will be ignored,
unless stated otherwise.

\subsubsection*{\uwave{Bremsstrahlung in the distortion spectrum.}\/}~
It can be seen from Fig.\,\ref{fig:dev1}, that for the most
realistic clusters with radii $R_{\rm c}=250-350$ kpc the
bremsstrahlung dominates over the Compton distortions in the
wide frequency range $20\ \mbox{MHz}\la \nu\la 3.5\ \mbox{GHz}.$
As the compactness of the cluster increases and as the density
of its intergalactic gas rises, this range extends to lower
frequencies. Obviously, a similar, but less pronounced effect is
also to be expected with increasing Thomson optical depth of the
cluster $\tau_{\rm T}$ at a constant radius $R_{\rm c}$ (in this
case, the Compton distortions also grow, but not so dramatically
as the bremsstrahlung intensity).

The gas bremsstrahlung contributes significantly to the range of
high frequencies of the background spectrum, completely or
partially compensating for the drop in the CMB flux related to
its Compton scattering by high-temperature electrons. To better
investigate this question, Fig.\,\ref{fig:devnu} presents the
distortions in the spectrum $\nu\,F_{\nu}(\nu)$ (the intensity
multiplied by the frequency) for a cluster with the same gas
temperature as that for the cluster in Fig.\,\ref{fig:dev1}, but
with a greater optical depth along the line of sight, $\tau_{\rm
  T}=1.2\times 10^{-2}$. We see that the bremsstrahlung
suppresses the drop in CMB brightness up to a frequency $\nu\sim
5$ GHz ($\lambda\sim 6$ cm) and gives rise to a radio source
toward the cluster on the maps of background fluctuations at
lower frequencies. Without the bremsstrahlung the ``shadow'' on
the background map would be observed up to a frequency
$\nu_2\simeq 802$ MHz. The bremsstrahlung also makes a
noticeable contribution at frequencies $\nu\ga 5$ GHz, reducing
the expected drop in the CMB flux in the Rayleigh-Jeans part of
the spectrum. This process should be taken into account when
interpreting the current (for example, VLA and BIMA, Dawson et
al. 2002) and planned (the Simons Observatory (SO), Ade et
al. 2019; CMB-S4, Abazajian et al. 2019; and CMB-HD, Sehgal et
al. 2019) active observations of the effect of a decrease in the
CMB brightness at centimeter wavelengths.

\begin{figure*}[tp]
\begin{minipage}{0.50\textwidth}
\includegraphics[width=1.0\linewidth]{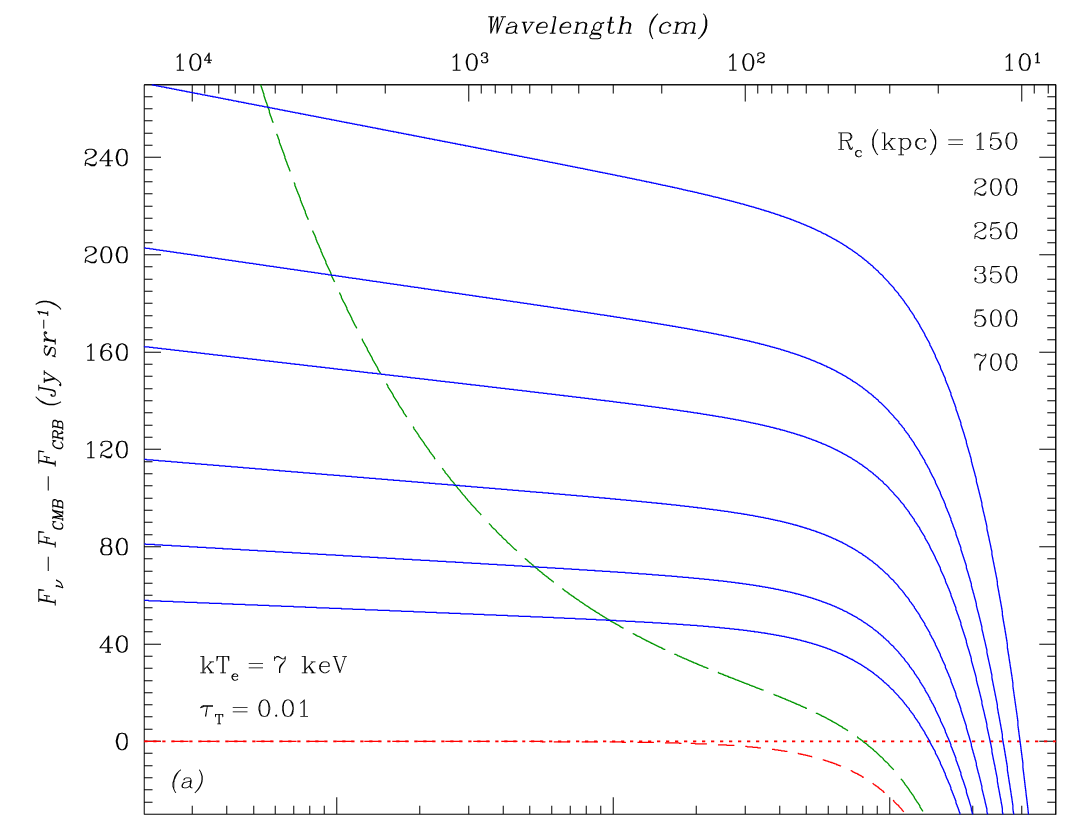}
\end{minipage}\hspace{-6mm}\begin{minipage}{0.50\textwidth}
\includegraphics[width=1.0\linewidth]{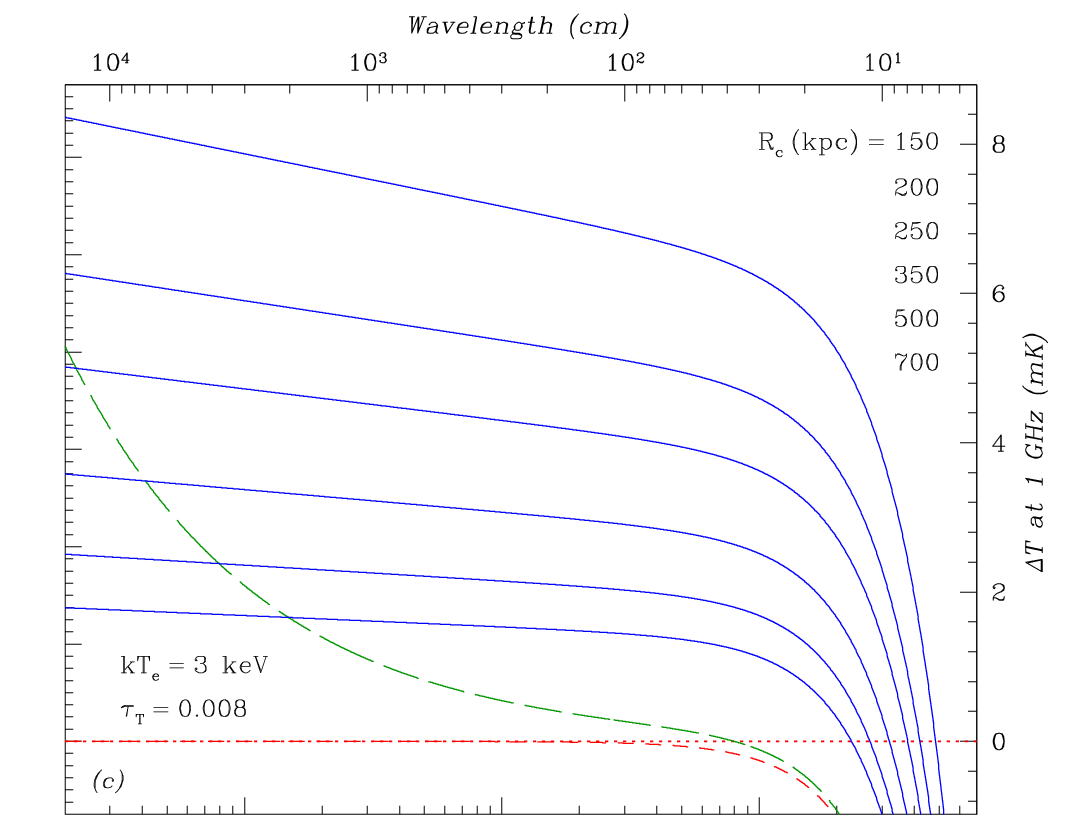}
\end{minipage}\\ [-2mm]
\begin{minipage}{0.50\textwidth}
\includegraphics[width=1.0\linewidth]{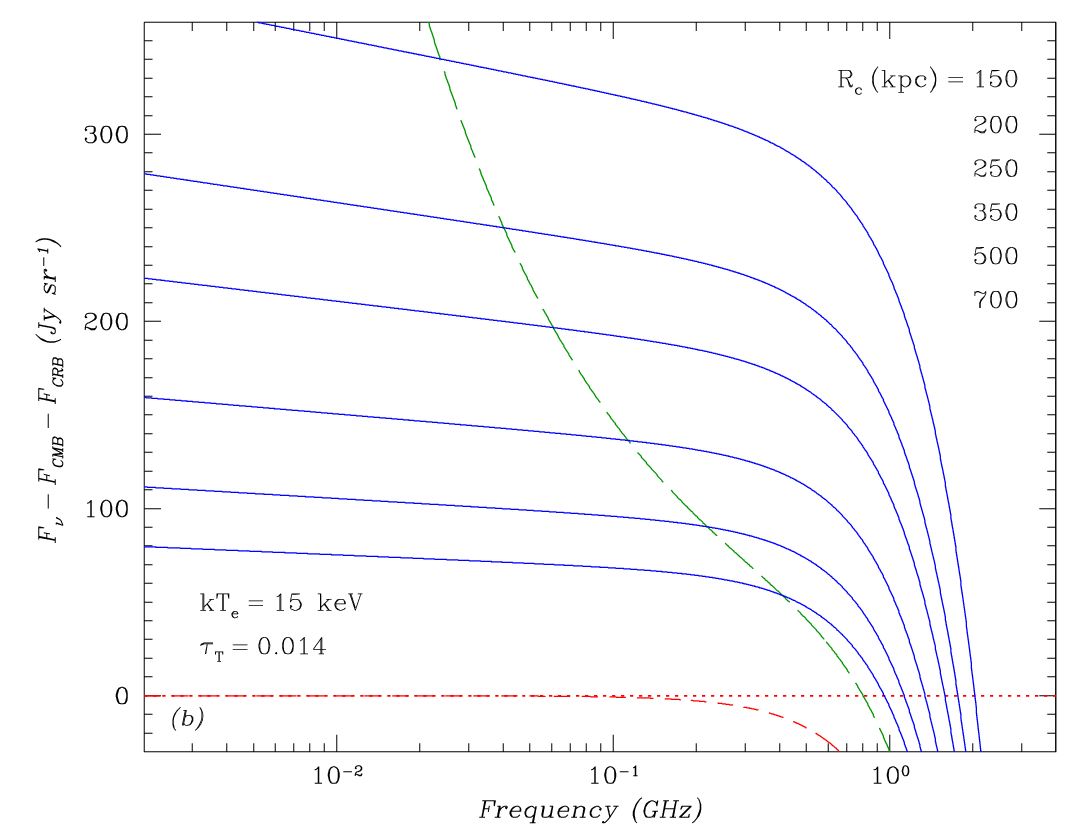}
\end{minipage}\hspace{-6mm}\begin{minipage}{0.50\textwidth}
\includegraphics[width=1.0\linewidth]{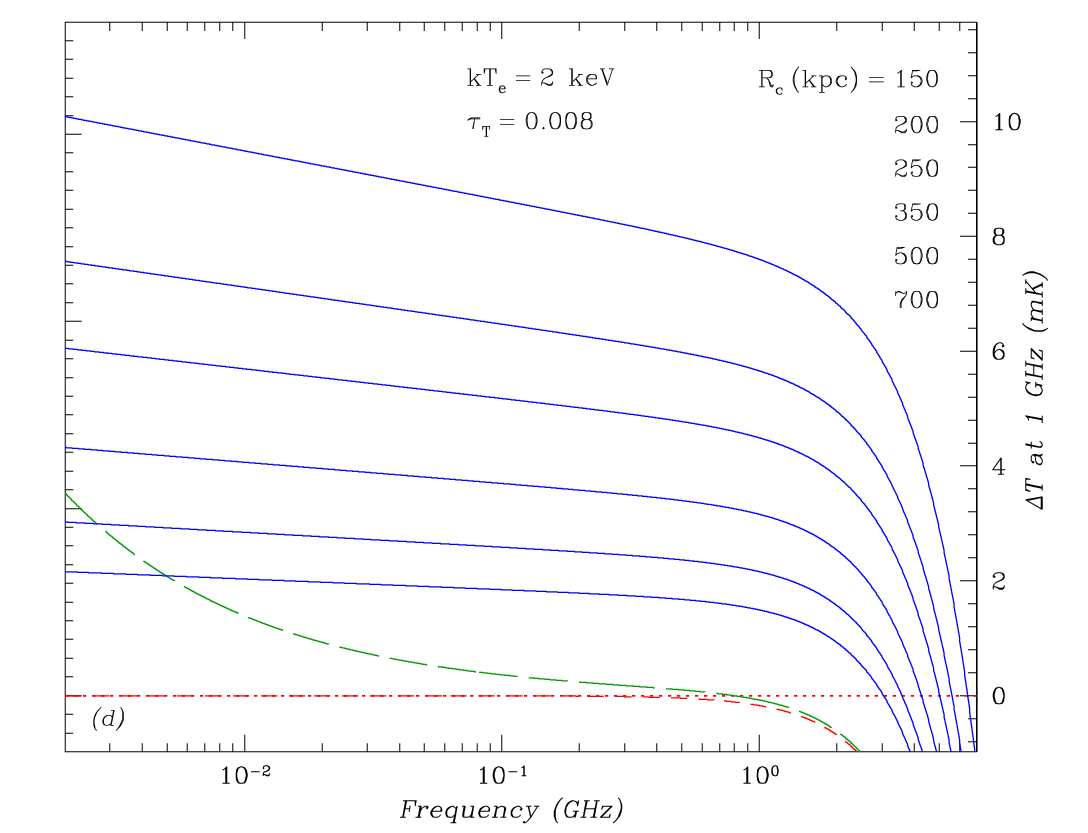}
\end{minipage}
\caption{\rm Same as Fig.\,\ref{fig:dev1}, but for clusters with
  the following parameters: {\sl (a)} $kT_{\rm e}=7$ keV,
  $\tau_{\rm T}=1\times10^{-2},$ {\sl (b)} $kT_{\rm e}=15$ keV,
  $\tau_{\rm T}=1.4\times10^{-2},$ {\sl (c)} $kT_{\rm e}=3$ keV,
  $\tau_{\rm T}=8\times10^{-3}$, and {\sl (d)} $kT_{\rm e}=2$ keV,
  $\tau_{\rm T}=8\times10^{-3}$. The bremsstrahlung (blue solid
  lines) dominates in a wider spectral range for cold, greatly
  relaxed clusters. At the same time, it completely compensates
  for the drop in CMB brightness up to frequencies of 5-8 GHz
  ($\lambda\simeq 4-6$ cm) and weakens it at higher
  frequencies. Although the distortions in the radio background
  spectrum become stronger for hot young clusters with a large
  optical depth, they become equal to the bremsstrahlung only at
  very low frequencies $\nu\la 20$ MHz ($\lambda\ga 15$ m).
  \label{fig:dev2} }
\end{figure*}

Figure~\ref{fig:dev2} shows dependences analogous to those
presented in Fig.\,\ref{fig:dev1}, but for clusters with
different parameters of the intergalactic gas cloud:
(Fig.\,\ref{fig:dev2}a) $kT_{\rm e}=7$ keV, $\tau_{\rm
  T}=1\times10^{-2};$ (Fig.\,\ref{fig:dev2}b) $kT_{\rm e}=15$
keV, $\tau_{\rm T}=1.4\times10^{-2};$ (Fig.\,\ref{fig:dev2}c)
$kT_{\rm e}=3$ keV, $\tau_{\rm T}=8\times10^{-3};$ and
(Fig.\,\ref{fig:dev2}d) $kT_{\rm e}=2$~keV, $\tau_{\rm
  T}=8\times10^{-3}$. The parameters of the sample clusters, the
temperature of the gas $kT_{\rm e}$, and its optical depth for
Thomson scattering $\tau_{\rm T}$, span a wide range of
values. It can be seen that in the overall spectrum of the radio
background distortions for cold (greatly relaxed) clusters the
gas bremsstrahlung dominates in a wider frequency range than in
the spectrum of hot young clusters. This is because of both the
increase in the intensity of the bremsstrahlung itself and the
decrease in the amplitude of the distortions in the radio
background spectrum upon Compton scattering by electrons
linearly dependent on their temperature. The bremsstrahlung
of cold clusters with $kT_{\rm e}=2-3$ keV also extends
much farther to high frequency (up to $\nu\sim 7-8$ GHz,
as can be seen even without constructing a figure
analogous to Fig.\,\ref{fig:devnu}). As noted above, here it
competes with the effect of a decrease in the CMB brightness due
to the Compton scattering by electrons. 

At the same time, the figure suggests that even for the hottest
sample clusters with $kT_{\rm e}=7-15$ keV there exists a
noticeable frequency range, $100\ \mbox{MHz}\la \nu\la
2.5\ \mbox{GHz},$ in which precisely the intergalactic gas
bremsstrahlung must be detected toward the cluster.  This is
important, since it implies that the transition from the
``shadow'' on the millimeter-centimeter background map toward the galaxy
cluster to the powerful ``radio source'' on the background map
at decimeter and meter wavelengths even in such rich young
clusters is associated precisely with the bremsstrahlung of
their hot intergalactic gas and not with the Compton distortions
of the radio background due to the scattering by gas electrons,
as Holder and Chluba (2021) assumed.
\begin{figure}[t]
\hspace{-4mm}\includegraphics[width=1.08\linewidth]{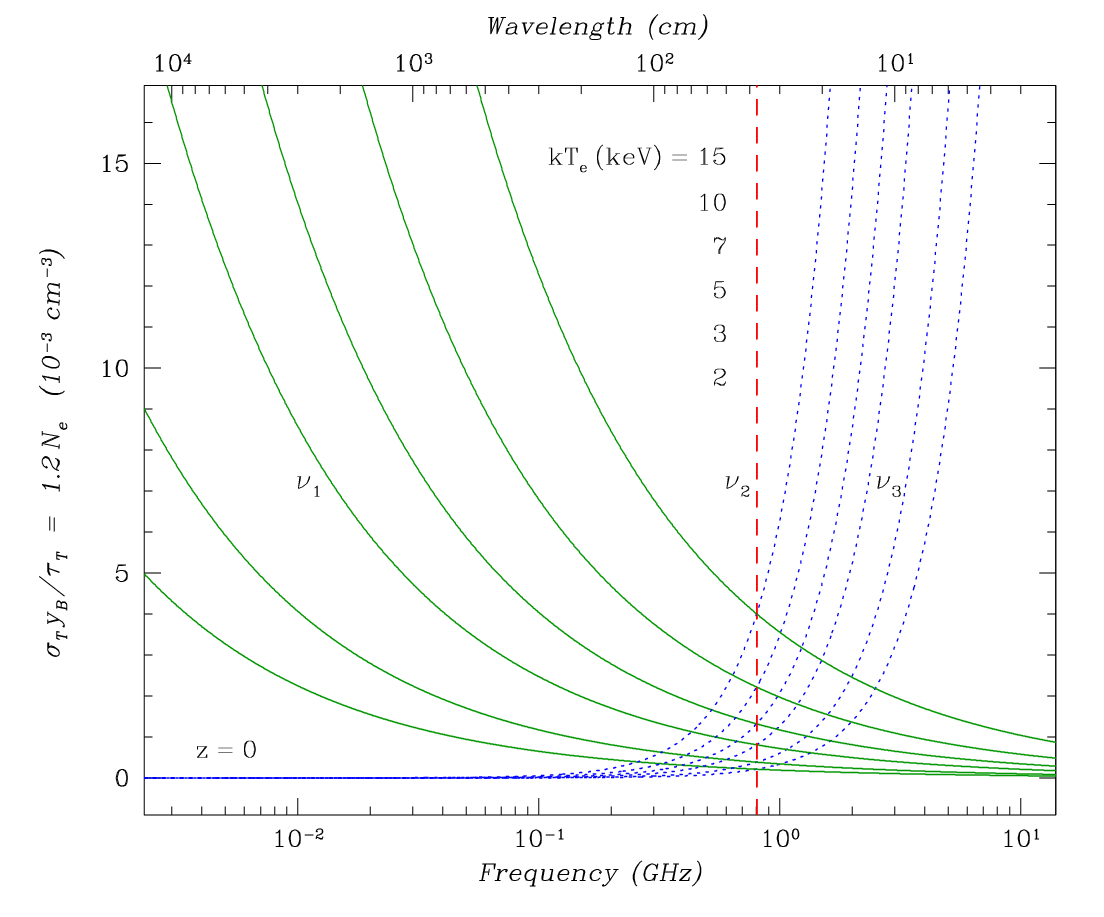}

\vspace{-1.5mm}
\caption{\rm The boundary frequency $\nu_1$ above which the
  bremsstrahlung flux from the intergalactic cluster gas
  dominates over the Compton enhancement of the radio background
  in the overall distortion spectrum (green solid curves) and
  frequency $\nu_3$ below which the bremsstrahlung dominates in
  absolute value over the Compton attenuation of the CMB flux
  (blue dotted curves). The curves for different gas
  temperatures in the cluster are shown. The vertical red dashed
  line indicates the frequency $\nu_2$ at which the Compton
  distortions of the radio background and the CMB become
  equal. The positions of the frequencies $\nu_1$ and $\nu_3$
  are given as a function of intergalactic gas temperature
  $kT_{\rm e}$ and density $N_{\rm e}$ (to be more precise,
  $\sigma_{\rm T}\,y_{\rm B}\,/y_{\rm C} \ [kT_{\rm e}/m_{\rm
      e}c^2]\,\simeq 1.18\,N_{\rm e}$).\label{fig:rmeq}}
\end{figure}

\subsubsection*{\uwave{Dependence on gas parameters.}\/}~Obviously,
the frequency range $\nu_1\la\nu\la\nu_3,$ where the radio
bremsstrahlung of the intergalactic gas dominates in the cluster
radiation, serves as the most important characteristic of a
galaxy cluster from the viewpoint of its detection on the
background maps as a radio source and the identification of the
physical process responsible for its appearance. The lower
boundary of this range $\nu_1$ is the frequency leftward of
which the enhancement of the radio background due to the Compton
scattering by electrons of the hot cluster gas exceeds its
bremsstrahlung flux. The upper boundary $\nu_3$ is the frequency
rightward of which the gas bremsstrahlung flux can no longer
compensate for the decrease in the brightness of the microwave
background. We explained how to calculate these critical
frequencies above in the paragraph following
Eq.~(\ref{eq:brem2}).

Figure~\ref{fig:rmeq} presents the results of our attempt to
investigate the positions of the frequencies $\nu_1$ and $\nu_3$
(and the width of the interval between them) as a function of
the main parameters of the hot cluster gas. Our analysis showed
that the electron temperature $kT_{\rm e}$ and the combination
of quantities $\sigma_{\rm T}\,y_{\rm B}\,/\tau_{\rm T},$ which
is equal to the electron number density in the cluster gas to
within a factor of $\simeq 1.18,$ are the optimal
independent parameters in this problem. This quantity is plotted
along the {\it Y\/} axis in the figure.  The solid and dotted
lines indicate the values of the frequencies $\nu_1$ and
$\nu_3$, respectively. Different curves, but of the same type,
correspond to different $kT_{\rm e}$. Obviously, the $\nu_1$ and
$\nu_3$ curves corresponding to the same temperature intersect
at the frequency $\nu_2$ for which in the absence of
bremsstrahlung the Compton distortions in the synchrotron
(power-law) radio background and the Planck microwave background
spectrum become equal in absolute value.

Figure~\ref{fig:rmeq} suggests that the region of frequencies
with the dominant contribution of the bremsstrahlung must be
present in the background distortion spectra toward all typical
nearby clusters. However, it narrows noticeably for very hot and
anomalously sparse clusters. The critical value of $(\sigma_{\rm
  T}\,y_{\rm B}\,/\tau_{\rm T})_{\rm min}\simeq 1.18\,N_{\rm
  e,min}$ at which this region should have collapsed
($\nu_1\rightarrow\nu_2,\ \nu_3\rightarrow\nu_2$) is shown in
Fig.\,\ref{fig:dmin} as a function of temperature $kT_{\rm e}$
and cluster redshift $z$ (see below). It can be seen that the
minimum gas density at which the bremsstrahlung region still
exists in the distortion spectrum increases with gas
temperature.

For clarity, on the right {\it Y\/} axis of
Fig.~\ref{fig:dmin} we show what the cluster radius $R_{\rm
  c}$ with the critical electron density must be. The Thomson
optical depth toward the cluster center is fixed at a
comparatively low value of $\tau_{\rm T}=0.006$ (barely sufficient
for the formation of a Compton distortion in the radio
background accessible to observations). Nevertheless, even in
the case of the hottest (but nearby, $z\simeq 0$) clusters in
this figure, with $kT_{\rm e}\ga 20$ keV, the gas bremsstrahlung
does not contribute to the radio flux only when their radius
$R_{\rm c}$ exceeds noticeably the actually observed cluster
sizes ($R_{\rm c}\ga 500$ kpc).
\begin{figure}[t]
\vspace{4.1mm}
  
\hspace{-3.5mm}\includegraphics[width=1.12\linewidth]{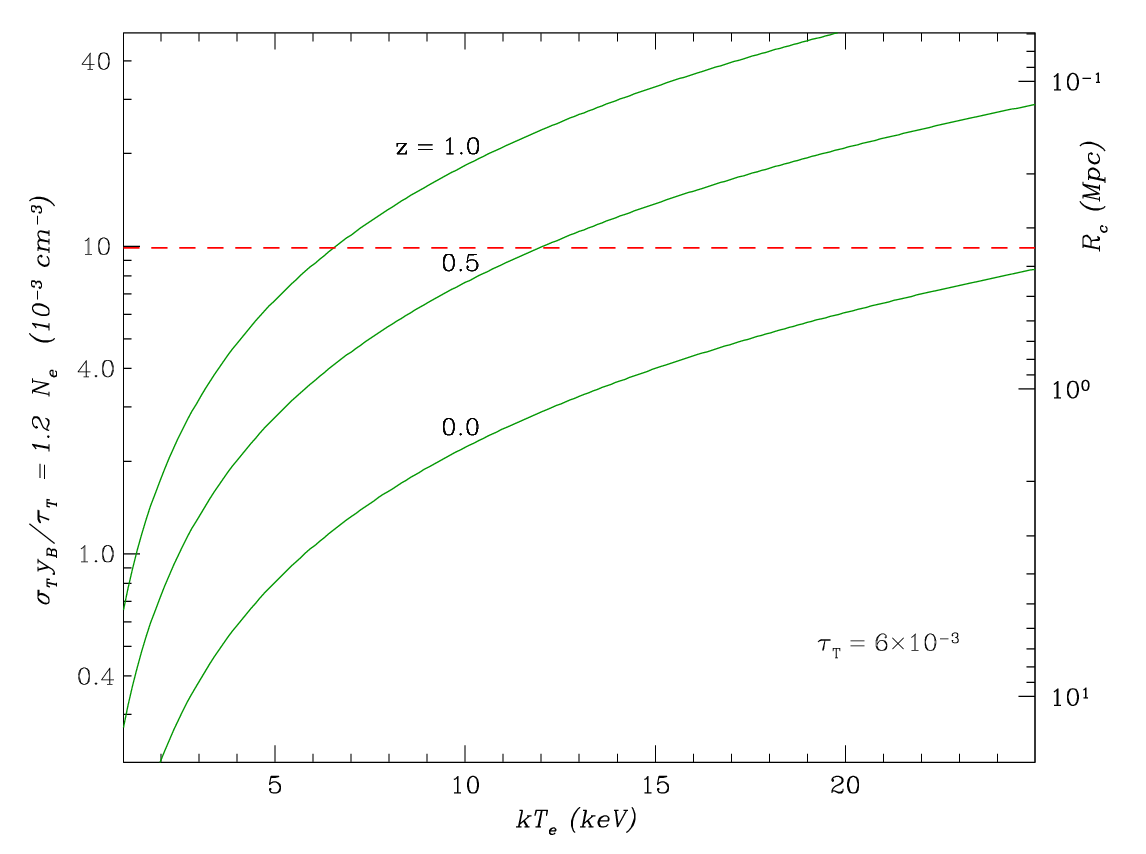}

\caption{\rm The minimum electron density $N_{\rm e,min}$ (to be
  more precise, $\sigma_{\rm T}\,y_{\rm B}\,/\tau_{\rm T}\simeq
  1.18\,N_{\rm e,min}$) of the hot intergalactic gas needed for
  the frequency range with the dominant contribution of the
  bremsstrahlung to exist in the spectrum of the radio
  background distortions toward the cluster. The density is
  higher for distant ($z>0$) clusters. The radius $R_{\rm c}$ of
  a cluster with the critical density and a Thomson optical
  depth toward the center $\tau_{\rm T}=6\times 10^{-3}$ is
  plotted on the {\it Y\/} axis on the right. The dashed line
  marks a typical cluster radius $R_{\rm c}\simeq 350$
  kpc.\label{fig:dmin}}
\end{figure}
\begin{figure}[!t]
\hspace{-4.5mm}\begin{minipage}{1.02\linewidth}
  \includegraphics[width=1.07\linewidth]{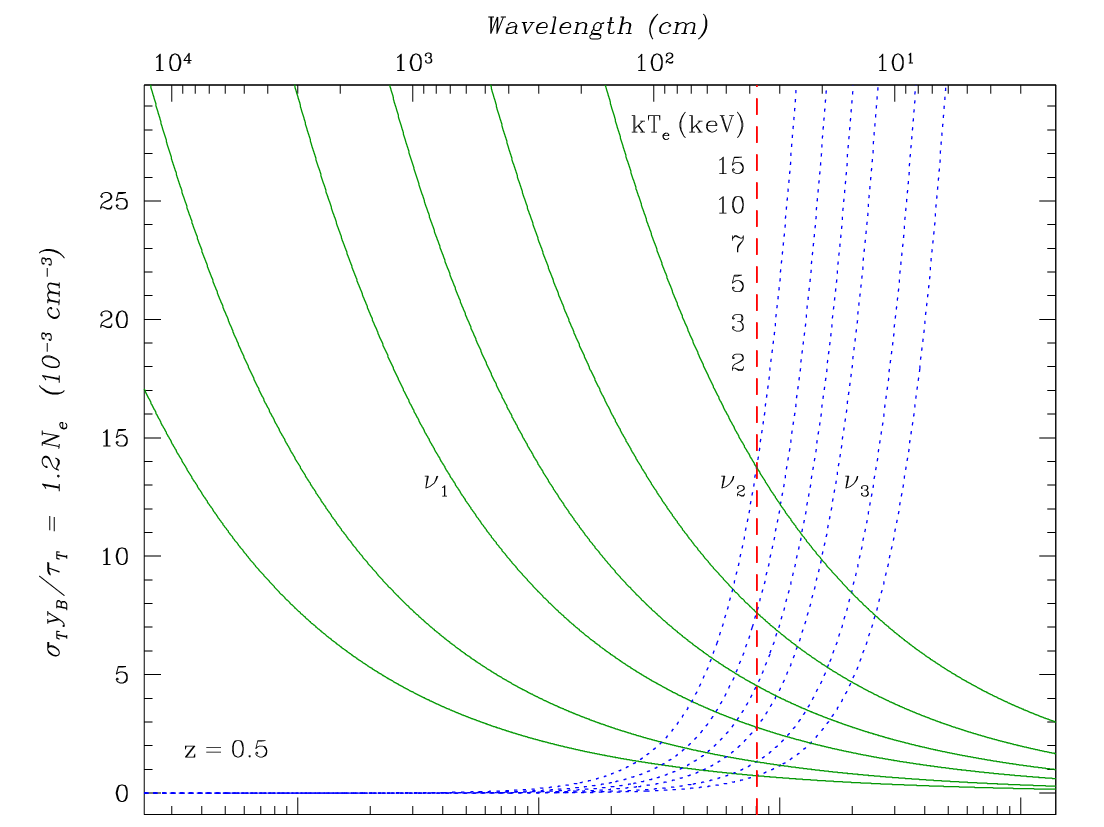}\\ [-2mm]
  \includegraphics[width=1.07\linewidth]{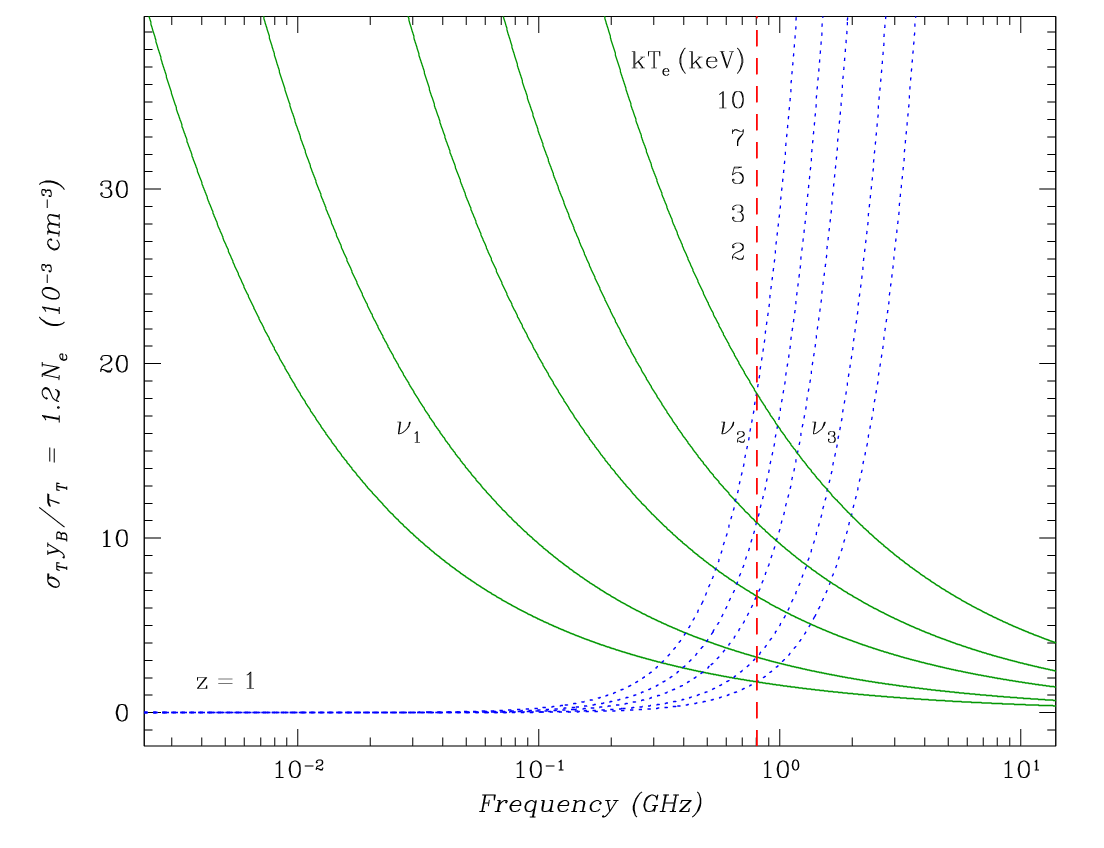}
  \end{minipage}
\caption{\rm Same as Fig.~\ref{fig:rmeq}, but for distant
  clusters at redshifts $z=0.5\ \mbox{and}\ 1$. Because of the
  drop in bremsstrahlung brightness with $z$, the frequency
  range in which the bremsstrahlung dominates in the background
  distortion spectrum toward the cluster narrows
  noticeably.\label{fig:rmez}}
\end{figure}

\subsubsection*{\uwave{Dependence on redshift $z$.}\/}~The
change in the spectrum of the microwave background with redshift
is completely determined by the dependence of its temperature on
$z$: $T_{\rm m}(z)=T_{\rm m}(1+z).$ The radio background
spectrum (Eq.~[\ref{eq:tcrb}]) depends on $z$ in a more
complicated way: $F_{R}(z)=
F_0\,(\nu/\nu_0)^{-\alpha}\,(1+z)^{3-\alpha}$ (Zeldovich
and Novikov 1975), and only under the assumption that the
power-law shape of the spectrum does not change with
redshift. We are interested in the background spectrum at the
present epoch, and, as has already been noted, it was measured
with a high accuracy. Its dependence on $z$ does not change the
distortions observed at our epoch that arise in it when
interacting with the intergalactic gas of galaxy clusters, no
matter how far they are.

The radio bremsstrahlung spectrum of this gas, whose brightness
changes as $F_{B}(z=0)=F_{B}(z)/(1+z)^3$ when observing
distant clusters (the gas temperature after its measurement was
assumed to have been corrected for $z$, i.e., reduced to the
cluster frame), is a different matter. Since the radio
bremsstrahlung spectrum itself depends weakly on the frequency,
$F_{B}\sim \nu^{-0.04}$, the change with $z$ is reduced to a
drop in the flux from distant clusters by the same factor at all
frequencies $(1+z)^{3}$. Figure\,\ref{fig:rmez} shows how the
frequency range $\nu_1<\nu< \nu_3$ with the dominant
bremsstrahlung in the distortion spectrum narrows when observing
clusters at redshifts $z=0.5\ \mbox{and}\ 1$. This figure is
analogous to Fig.\,\ref{fig:rmeq}, which presents the same
frequency range $\nu_1<\nu< \nu_3$, but typical for local ($z\ll
1$) clusters.

Figure~\ref{fig:dmin} presents the curves of the critical
density $N_{\rm e,min}$ at which the frequency range with the
dominant bremsstrahlung is still absent in the distortion
spectrum for two such distant ($z=0.5\ \mbox{and}\ 1$)
clusters. It can be seen that as the redshift of the cluster
increases, the observed intensity of its radio bremsstrahlung
drops rapidly; accordingly, it may not be detected at all
against the Compton radio background distortions even in the
case when it has a lower temperature and a higher gas density
than do similar local clusters. The constraints on the critical
cluster radius also weakens.

\section*{REAL CLUSTERS}
\noindent
Above we have assumed that the radio flux is measured toward the
cluster center and with a good angular resolution (compared to
the angular size of the cluster). Obviously, the peripheral
observations of the gas in clusters with a uniform density
distribution are less preferable, since in this case both the
gas bremsstrahlung flux and the amplitude of the radio background
distortions due to the scattering decrease equally
(proportionally to $l$ --- the cluster size along the line of
sight). The situation is different if the peripheral
observations are carried out in the case of a real cluster with
a density that falls off slowly with radius. Since the amplitude
of the background distortions due to the scattering and the
bremsstrahlung intensity are proportional to $N_{\rm e}$ and
$N_{\rm e}^2$, respectively, the contribution of the
bremsstrahlung to the observed radio background excess decreases
to the cluster edge faster than does the contribution of the
scattering (Zeldovich and Sunyaev 1982).

\subsubsection*{\uwave{Nonuniform density distribution.}\/}~Let
us illustrate this effect using a cluster characterized by a
$\beta$ gas density distribution along the radius (Cavaliere and
Fusco-Femiano 1976),
\begin{equation}\label{eq:king}
  N_{\rm e}=N_{\rm c}\left(1+\frac{R^2}{R_{\rm
      c}^2}\right)^{-3\beta/2},
\end{equation} with a central electron density $N_c$ and a
parameter $\beta\simeq 2/3$ that agrees well with the observed
X-ray brightness distribution of many galaxy clusters (Jones and
Forman 1984; Arnaud 2009) as an example. For such a cluster the
bremsstrahlung parameter (emission measure) of the intergalactic
gas $y_{\rm B}(\rho) = 2 \int_0^{\infty} \Sigma\ Z^2 N_{\rm
  Z}(R) N_{\rm e}(R)\,d\,l$, which defines the intensity of its
thermal radiation along the line of sight at an impact parameter
$\rho$ from the direction toward the center, is
\begin{equation}\label{eq:kingme}
  y_{\rm B}(\rho)=0.59\,\pi\left(1+\frac{\rho^2}{R_{\rm
      c}^2}\right)^{-3/2}\, N_{\rm c}^2\,R_{\rm c}.
\end{equation}
The radius $R$, the impact parameter $\rho$, and the distance
along the line of sight $l$ are related by the
expression $R^2=\rho^2+l^2$.

The Thomson optical depth of the gas along the line of sight
$\tau_{\rm T}(\rho)=2\sigma_{\rm T}\int_0^{\infty}N_{\rm
  e}(R)\,d\,l,$ which defines the amplitude of the spectral
distortions due to the scattering, is
\begin{equation}\label{eq:kingtau}
  \tau{\rm}_{\rm T}(\rho)=\pi\ \left(1+\frac{\rho^2}{R_{\rm
      c}^2}\right)^{-1/2}\, \sigma_{\rm T}\, N_{\rm c}\,R_{\rm c}.
\end{equation}
\begin{figure}[!t]
\hspace{-4.5mm}\includegraphics[width=1.08\linewidth]{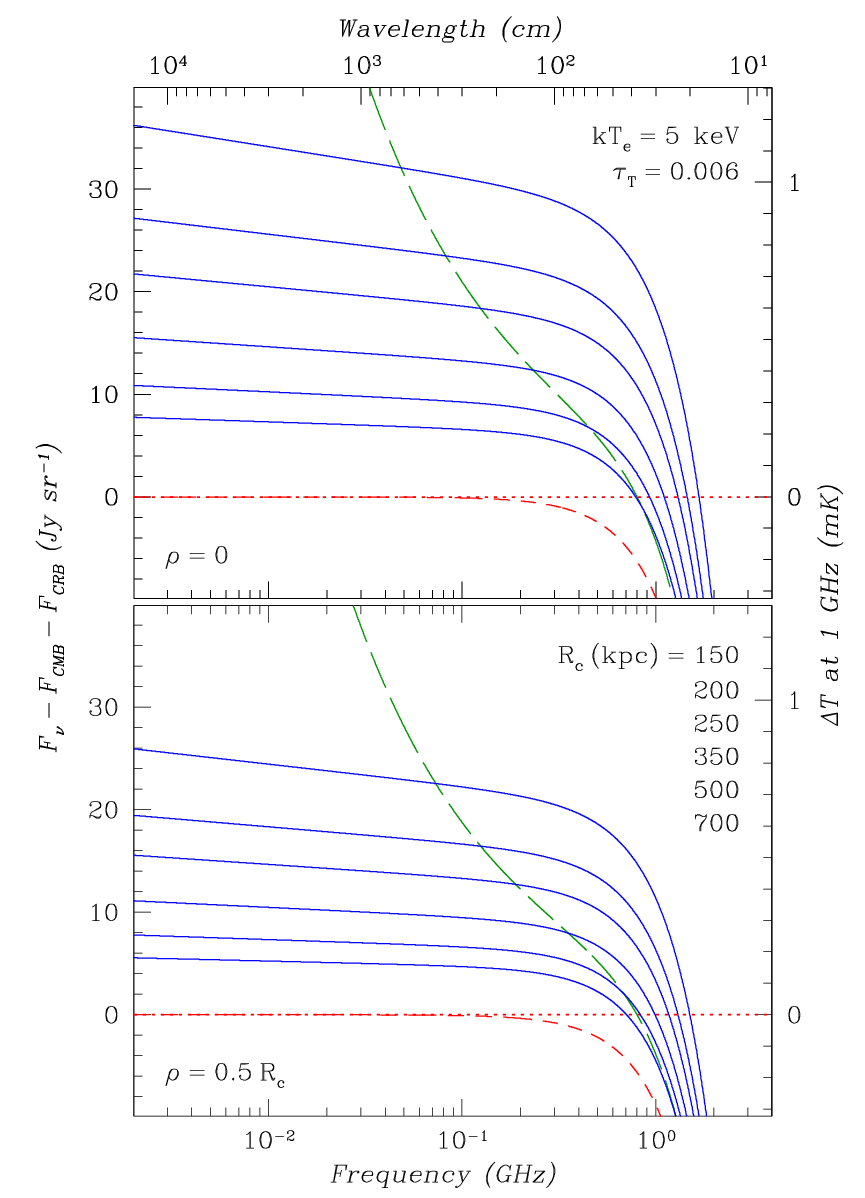}
\caption{\rm Same as Fig.~\ref{fig:dev1}, but for a cluster with
  a $\beta$ gas density distribution ($\beta=2/3$). The Thomson
  optical depth toward the cluster center is $\tau_{\rm
    T}=6\times 10^{-3}$, the gas is isothermal with $kT_{\rm
    e}=5$ keV, the cluster is a nearby one ($z\ll 1$). The
  contribution of the bremsstrahlung is indicated by the blue
  solid lines for different cluster core radii $R_{\rm c}$. The
  central (impact parameter $\rho=0$, {\it the upper panel}) and
  peripheral ($\rho=0.5\ R_{\rm c}$, {\it the lower panel}) observations
  are considered.\label{fig:beta-spec}}
\end{figure}
Figure~\ref{fig:beta-spec} presents the distortions in the radio
and microwave background spectrum in such a cluster expected due
to the Compton scattering by electrons (green dashes) and the
contribution of the thermal bremsstrahlung (blue solid lines) at
two impact parameters, $\rho=0$ and $0.5\ R_{\rm c}$. The gas
parameters, the radial Thomson optical depth $\tau_{\rm
  c}=3\times 10^{-3}$ and the electron temperature $kT_{\rm
  e}=5$ keV, were chosen to be the same as those for a cluster
with a uniform density distribution in
Fig.~\ref{fig:dev1}. Various values of the cluster core radius
$R_{\rm c}$ were considered. It can be seen that the Compton
background distortions toward the cluster center ($\rho=0$) in
Figs.~\ref{fig:dev1} and \ref{fig:beta-spec} are indeed the same
because of the equality of the optical depths. However, the
intensity of the thermal radiation from the gas of a real
cluster is a factor of $\pi$ lower than the intensity of the
radiation from a cluster with a uniform gas density
distribution.

This result is easy to explain: although the gas in a real
cluster is more strongly concentrated to the center, its density
is slightly lower (by a factor of $\pi/2\simeq1.57$) than the
gas density in a homogeneous cluster with the same core radius
$R_{\rm c}$ and optical depth $\tau_{\rm T}$. A sizeable
fraction of the optical depth is gained at large radii $R>R_{\rm
  c}$. Similarly, the cluster gas mass $$M_{\rm
  g}(<R)=4.6\left(\frac{\tau_{\rm T} m_{\rm p}}{\sigma_{\rm
    T}}\right) R_{\rm c}^2 \left[\frac{R}{R_{\rm
      c}}-\mbox{arctg}\left(\frac{R}{R_{\rm c}}\right)\right]$$
within the radius $R=R_{\rm c}$ is comparatively small, $M_{\rm
  g}(<R_{\rm c})\simeq 4.8 \times 10^{12}\ M_{\odot},$ but it
increases rapidly with $R$, reaching $M_{\rm g}(<2\ R_{\rm
  c})\simeq 2\times 10^{13}\ M_{\odot}$ at $R=2 R_{\rm c}$,
which is already close to the gas mass in a homogeneous cluster
($\simeq 2.1\times10^{13}\ M_{\odot}$ at $\tau_{\rm T}=0.06$).

The intensity of the thermal radiation from the gas of a real
cluster decreases still more dramatically when it is observed
even at a small ($\rho\sim 0.5\ R_{\rm c}$,
Fig.~\ref{fig:beta-spec}, {\it lower panel}) distance from the
direction toward the center.  In the case of such peripheral
observations, the action of the Compton scattering effect also
weakens with regard to both radio (green dashes) and microwave
(short red dashes) backgrounds (and equally, so that the
frequency $\nu_2$ does not change), but this occurs much more
slowly. Therefore, the peripheral observations of real clusters
turn out to be advantageous from the viewpoint of detecting the
scattering effect.
\begin{figure}[!t]
\vspace{-1mm}
\hspace{-4mm}\includegraphics[width=1.06\linewidth]{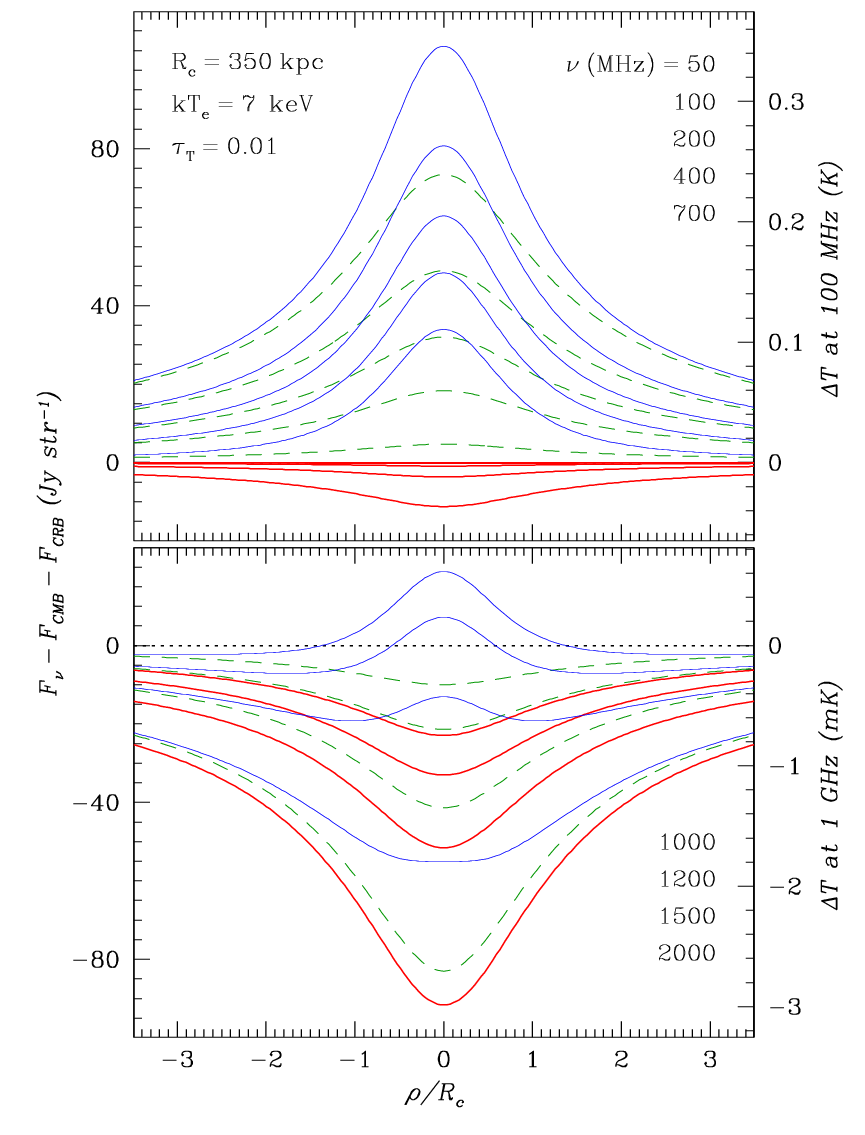}

\caption{\rm Comparison of the distortions in the radio
  background due to its scattering by electrons of the hot
  cluster gas (green dashes) and the bremsstrahlung of this gas
  (blue solid lines) at various frequencies as a function of the
  impact parameter $\rho/R_{\rm c}$ from the direction toward
  the cluster center. The CMB distortions due to the scattering
  by gas electrons (red solid lines) were also taken into
  account. The computation for a nearby ($z\ll 1$) cluster with
  a $\beta$ density distribution ($\beta=2/3$), a core radius
  $R_{\rm c} = 350$ kpc, a temperature $kT_{\rm e}=7$ keV, and a
  Thomson optical depth toward the center $\tau_{\rm T}=1\times
  10^{-2}$ (see Fig.\,\ref{fig:dev2}a).\label{fig:beta-flux}}
\end{figure}

Figure~\ref{fig:beta-flux}, which is also based on
Eqs.~(\ref{eq:kingme}) and(\ref{eq:kingtau}), shows how the
brightness of the cosmic radio background at various frequencies
changes with impact parameter $\rho$. We again considered a
nearby cluster with a $\beta$ density distribution, but with a
higher temperature $kT_{\rm e}=7$ keV and an optical depth
$\tau_{\rm T}=0.01$ of the gas (as in Fig.~\ref{fig:dev2}a). The
cluster core radius is $R_{\rm c}=350$ kpc. It can be seen that
the thermal radiation of the gas toward the cluster center
dominates in the increase in radio brightness at frequencies
$\ga 200$ MHz, but even at a small ($\rho\ga 2 R_{\rm c}$)
distance from this direction a major contribution to the
increase in the brightness of the radio background is made by
its scattering by electrons at all of the frequencies
considered. As the frequency increases, the region with the
dominant thermal radiation slightly expands.

Note that in Fig.~\ref{fig:beta-flux} (its upper panel) (at
frequencies $\la 700$ MHz) the thermal and scattered radio
emissions form a positive source at any $\rho$. Positive
radiation is observed on the lower panel only at small $\rho\la
R_{\rm c}$ and frequencies $\nu \la 1200$ MHz. The point is that
when calculating the distortions, in addition to the Compton
scattering of the radio emission, we took into account the
scattering of the CMB by electrons, which leads to a decrease in
background brightness in the central cluster regions (Sunyaev
and Zeldovich 1970, 1972; indicated by the red solid lines). As
mentioned above, the Compton distortions of the microwave and
radio emissions become equal in absolute value (compensate each
other) at $\nu_2\simeq 802$ MHz, and this can be seen on the
upper panel of Fig.~\ref{fig:beta-flux}. At higher frequencies
the observed background excess is associated exclusively with
the thermal gas radiation.

It can be seen from Fig.~\ref{fig:beta-flux} that (1) the
decrease in the brightness of the microwave background toward
the cluster center begins to be suppressed by the intrinsic
thermal radiation of the gas starting already at frequencies
$\nu\sim 2.0$ GHz, being completely compensated near $\nu\sim
1.3$ GHz, (2) even at small distances from the center $\rho\ga
R_{\rm c}$ the reduced background is retained up to frequencies
$\sim 800$ MHz, and (3) at frequencies $\nu\la 300$ MHz the
decrease in CMB brightness disappears completely irrespective of
the Compton scattering of the radio emission or the contribution
of the thermal gas radiation. This is because of the fast
natural (Rayleigh-Jeans) drop in the CMB flux when moving
downward along the frequency axis and the equally fast rise in
the flux of the cosmic radio background.
\begin{figure}[tp]
  \centering
\includegraphics[width=0.598\linewidth]{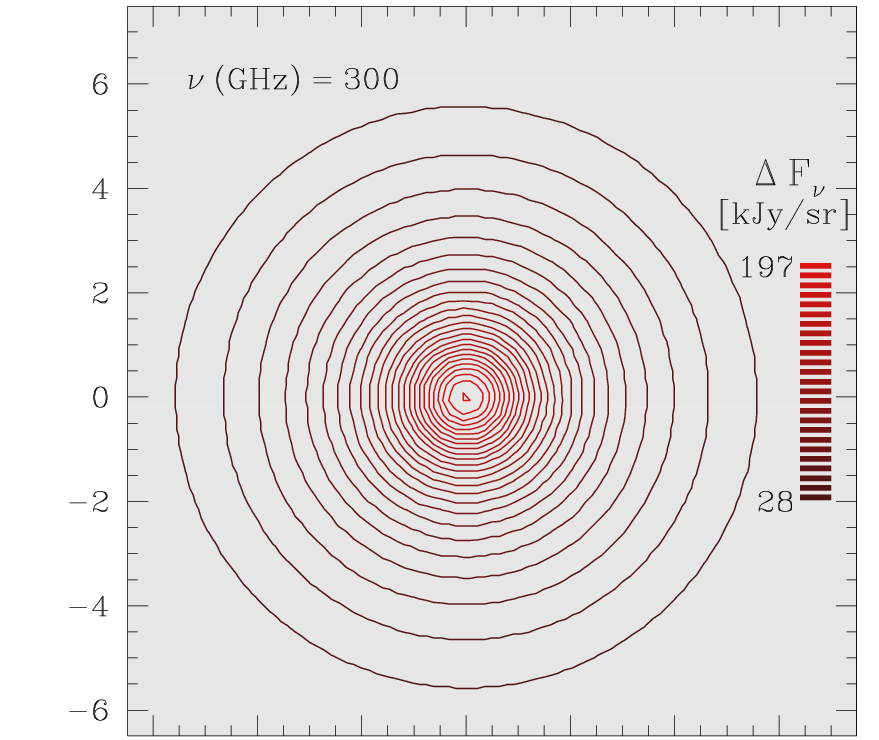}\\
\includegraphics[width=0.598\linewidth]{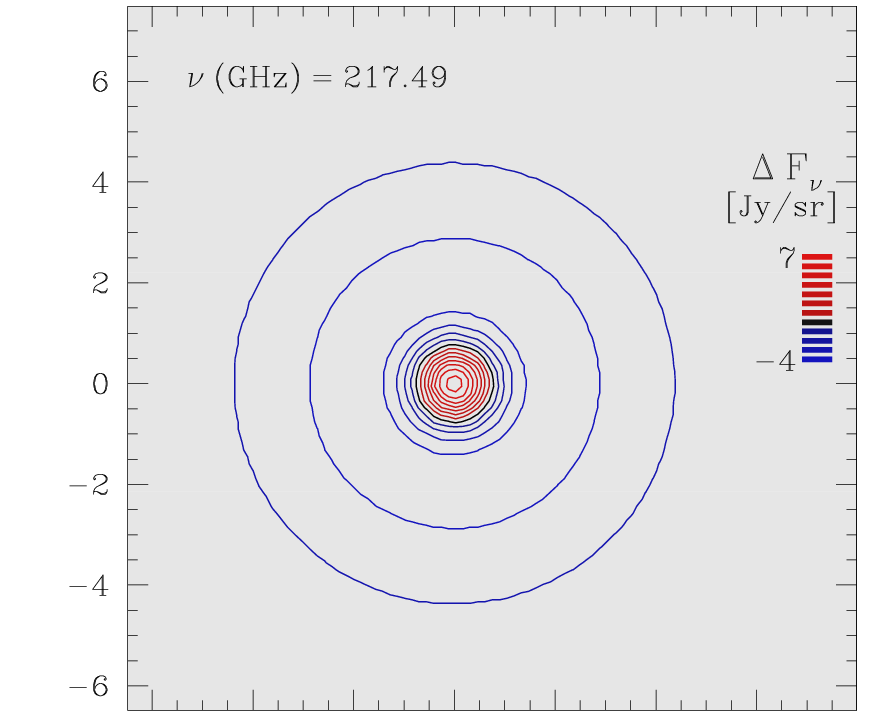}\\
\includegraphics[width=0.598\linewidth]{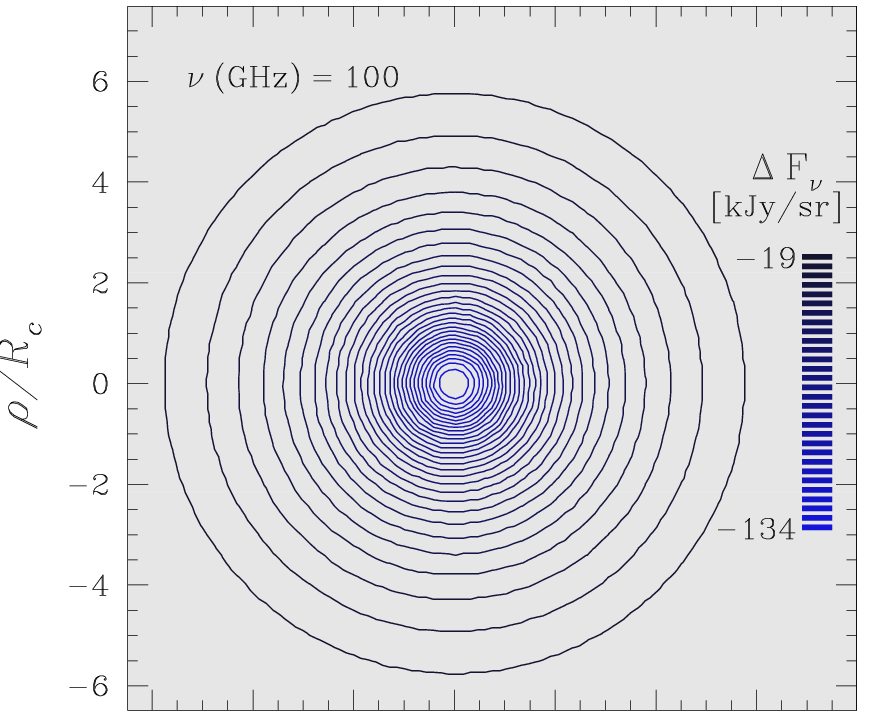}\\
\includegraphics[width=0.598\linewidth]{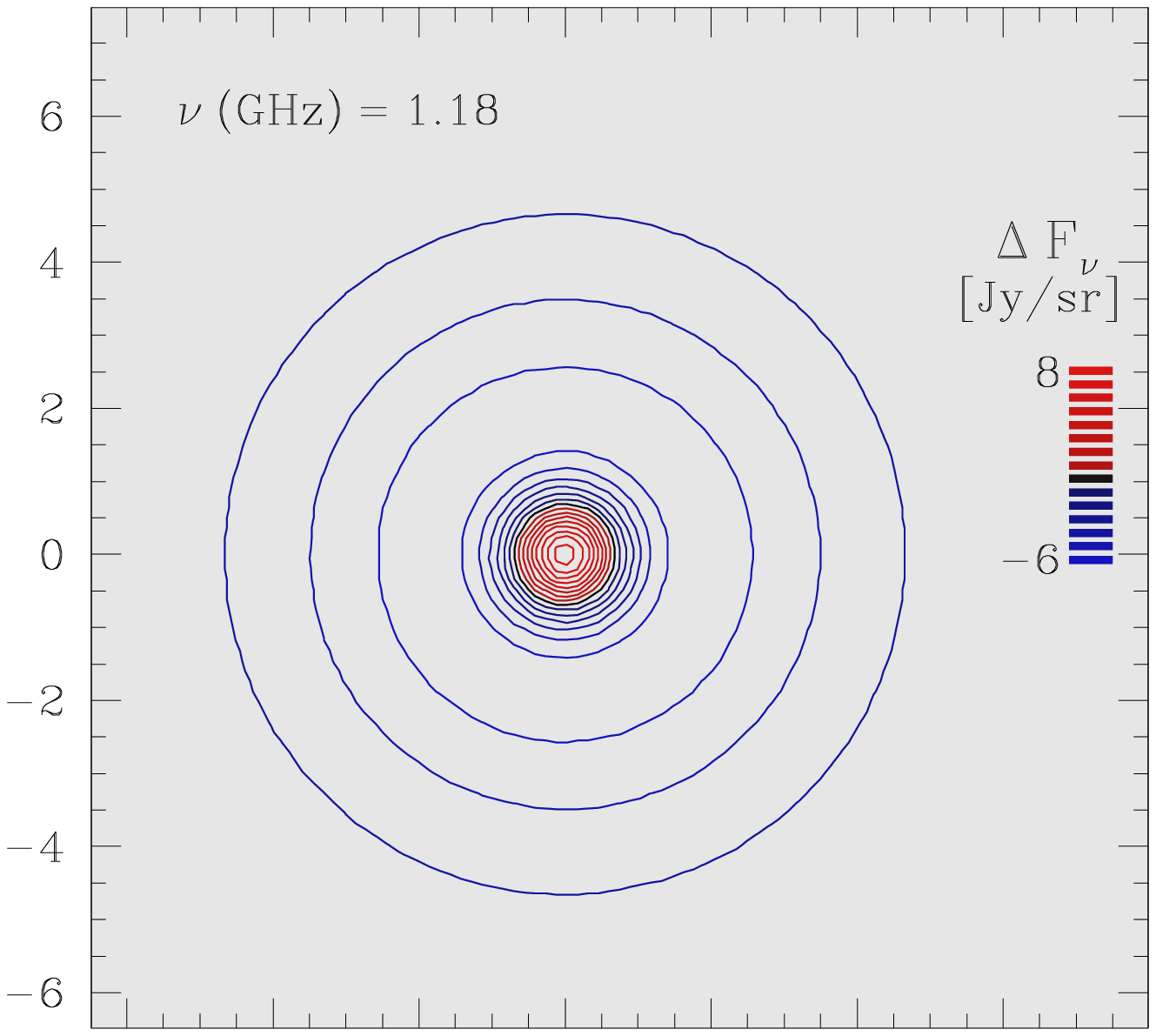}\\
\includegraphics[width=0.598\linewidth]{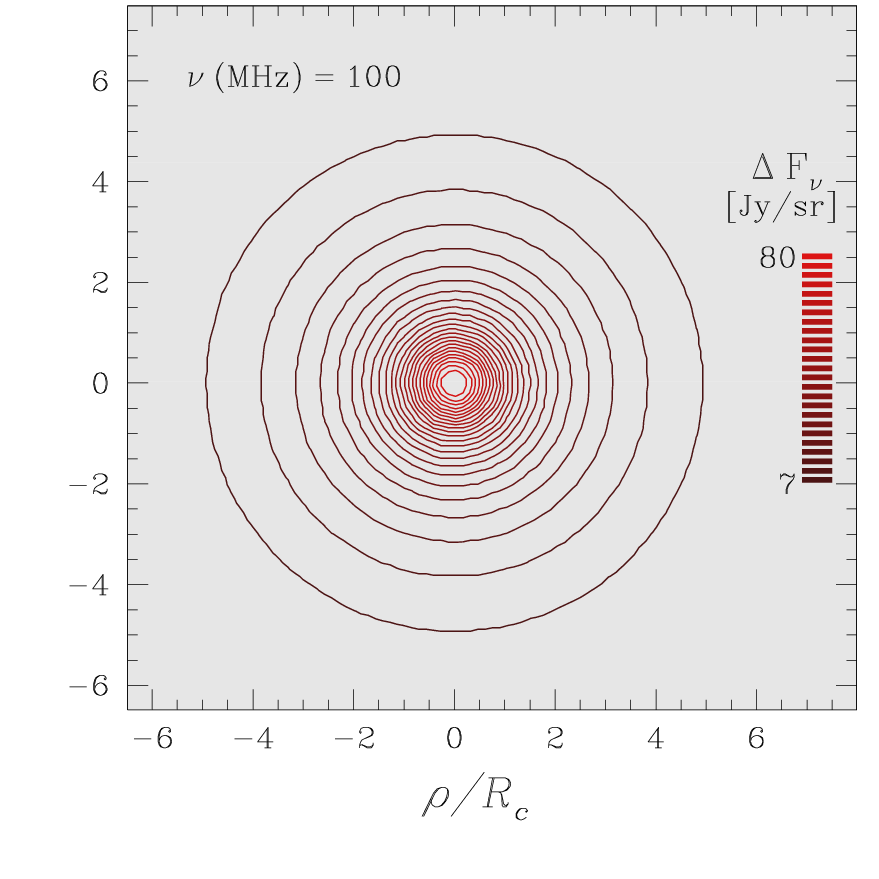}\\ [-5mm]

\caption{\rm Map of cosmic microwave and radio background
  distortions toward the same galaxy cluster as that in
  Fig.\,\ref{fig:beta-flux} at various frequencies (the blue and
  red lines indicate the ``negative'' and ``positive''
  deviations, respectively). The outer contours (at distances
  $\rho \ga 6\,R_{\rm c}$) are not shown.\label{fig:maps}}
\end{figure}

\subsubsection*{\uwave{A source of hybrid shape at the location of
the cluster.}\/}~Obviously, the downward transition in frequency
in the spectrum of the background distortions measured toward a
galaxy cluster from the region of decreased brightness (CMB) to
the region of increased brightness (radio emission) must be
accompanied by a change in the image of the source on the maps
of background fluctuations. Figure~\ref{fig:beta-flux} (its lower
panel) predicts that in the frequency range $1.0\ \mbox{\rm
  GHz}\la \nu\la 1.2$ GHz the cluster image must take a highly
unusual form --- a bright source at the center (excess
bremsstrahlung) surrounded by a ringlike shadow region (a region
of CMB deficit). This can be seen even better from
Fig.\,\ref{fig:maps} that shows the modeled maps of microwave
and radio background distortions at various frequencies toward a
nearby galaxy cluster with the same gas parameters as those used
in constructing Fig.\,\ref{fig:beta-flux}. The unperturbed
background was subtracted.

An unusual (hybrid) source is seen on the map for $\nu=1.18$
GHz. At higher frequencies ($\nu=100$ GHz) the central source
disappears and the shadow region expands to a circle with a
radius exceeding noticeably the cluster core
radius.\footnote{The shadow region in the hybrid source actually
  extends also noticeably farther along the radius than follows
  from the figure, but in this source it has a smaller depth. The
  low-intensity levels corresponding to the far wings of the
  source (large impact parameters $\rho/R_{\rm c}\ga 6$) are
  simply not shown on these panels.}. The source on the map of
background fluctuations at the location of the cluster becomes
``negative'' --- a ``hole'' in the background. It is known to
turn again into an ordinary ``positive'' source at frequencies
above $\nu_0 \simeq 217.5$ GHz ($\lambda\simeq 1.37$ mm, Sunyaev
and Zeldovich 1980, 1982) due to the lower-frequency CMB photons
thrown into this region by the inverse Compton scattering. This
is illustrated by the map for a frequency of 300 GHz.  At the
other edge of the spectrum, at frequencies $\la 800\ \mbox{\rm
  MHz}$, the hybrid source at the location of the cluster
described above also turns into a completely ``positive'' source
primarily due to the Rayleigh-Jeans decrease in the CMB flux and,
accordingly, in the absolute depth of the Compton ``hole'' forming
in it, but also due to the excess associated with the Compton
scattering of the growing radio background and the contribution
of the intergalactic gas bremsstrahlung. In Fig.~\ref{fig:maps}
the transition to the ``positive'' source is illustrated by the
lower map computed for $\nu= 100\ \mbox{\rm MHz}.$

The appearance of a source with a hybrid shape on the map of
background fluctuations is explained by the strong concentration
of the thermal (bremsstrahlung) radiation from the intergalactic
gas to the cluster center. The Compton CMB distortions forming
the ``shadow'' (``hole'') on the map are characterized by a
larger spatial scale, which is explained by their weaker
(linear) dependence on the hot-gas electron density.
\begin{figure*}[hp]
  \hspace{10mm}
  \begin{minipage}{1.00\textwidth}
  \includegraphics[height=0.31\linewidth]{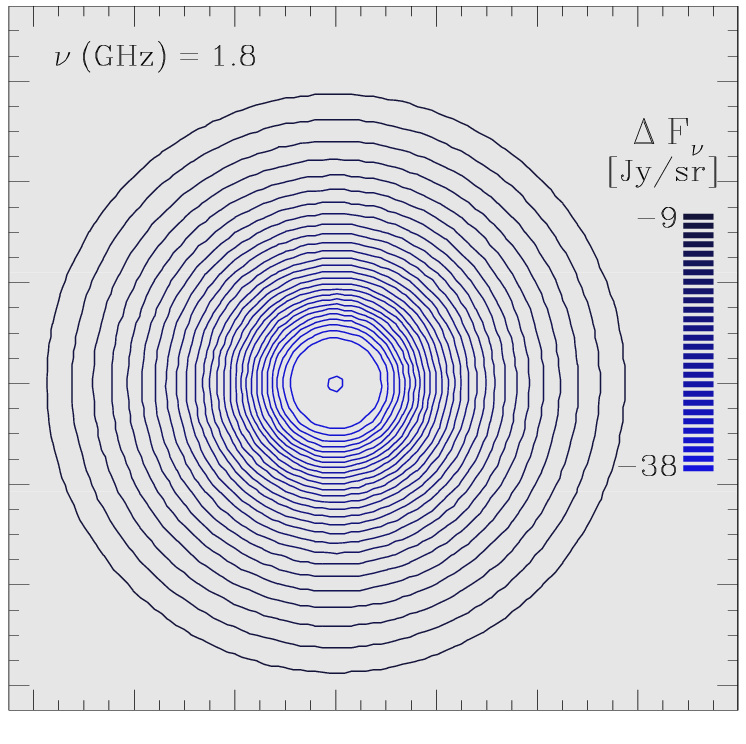}
  \hspace{-1.6mm}
  \includegraphics[height=0.31\linewidth]{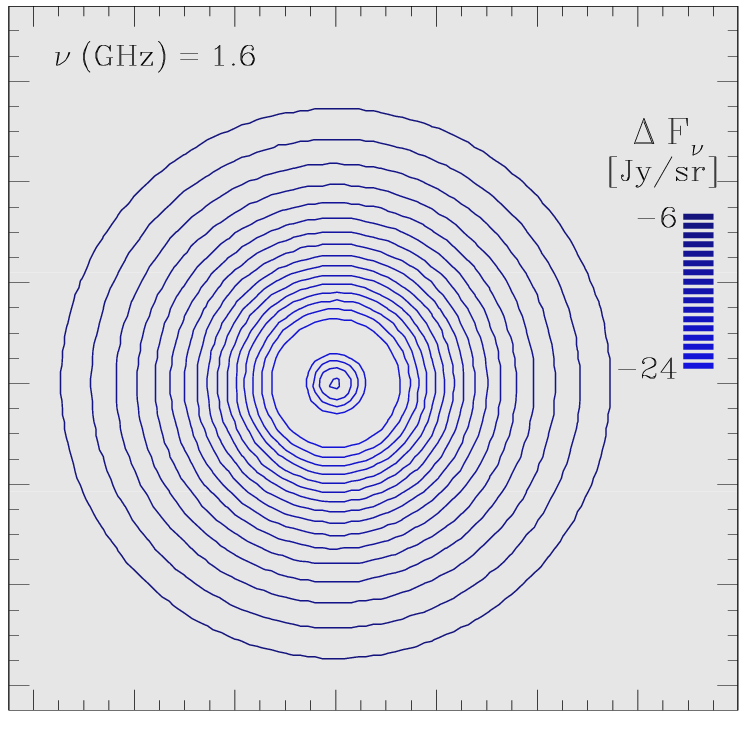}
  \hspace{-1.6mm}
  \includegraphics[height=0.31\linewidth]{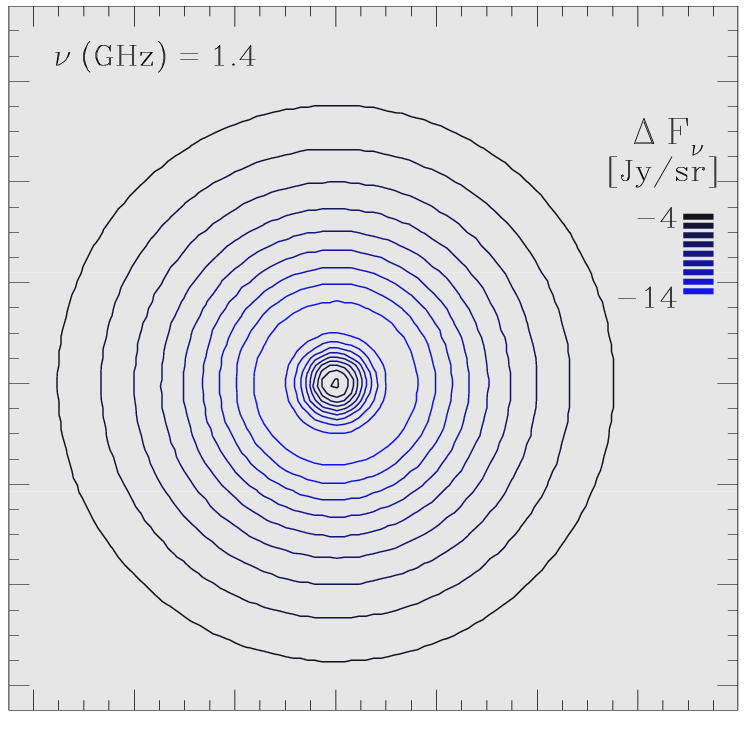}\\ [-1.7mm]
  \includegraphics[height=0.31\linewidth]{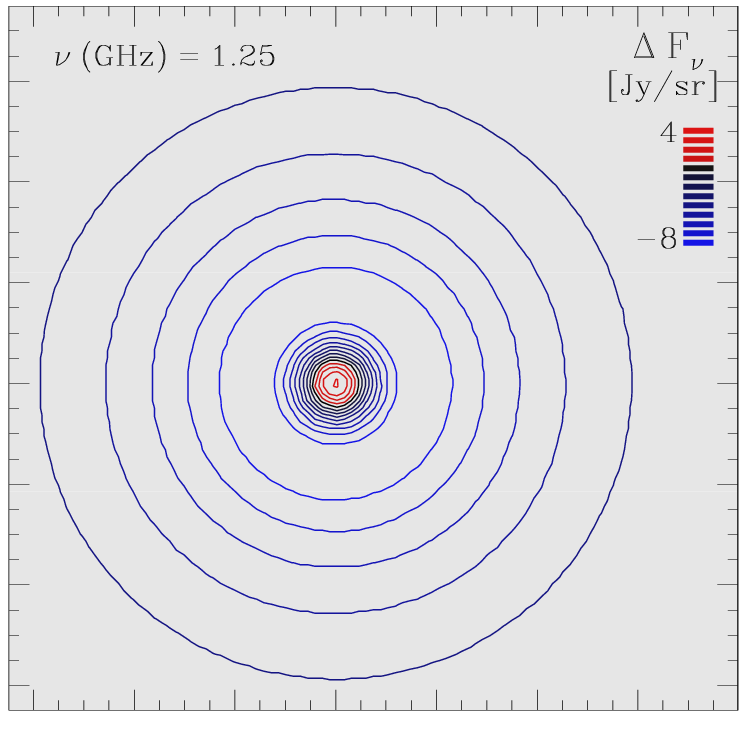}
  \hspace{-1.6mm}
  \includegraphics[height=0.31\linewidth]{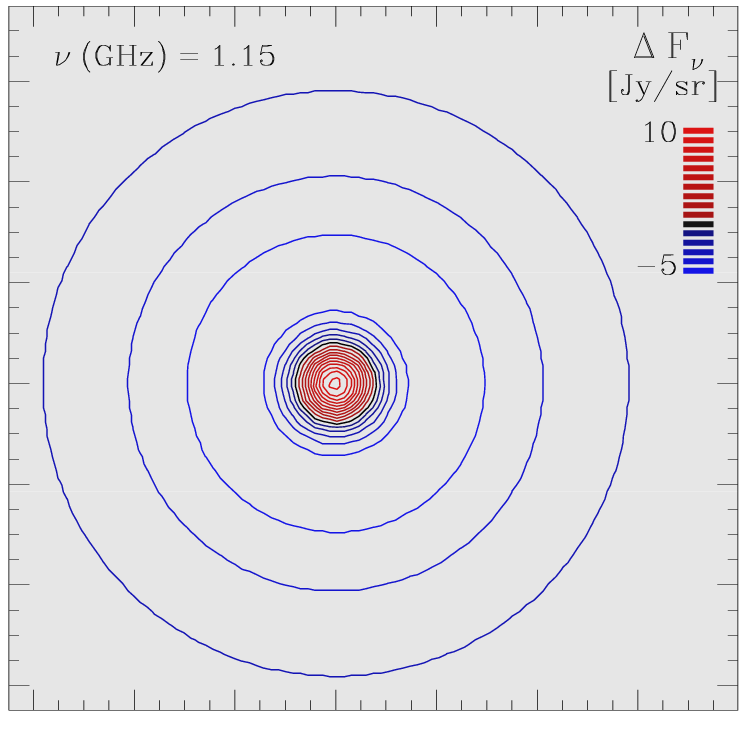}
  \hspace{-1.6mm} 
  \includegraphics[height=0.31\linewidth]{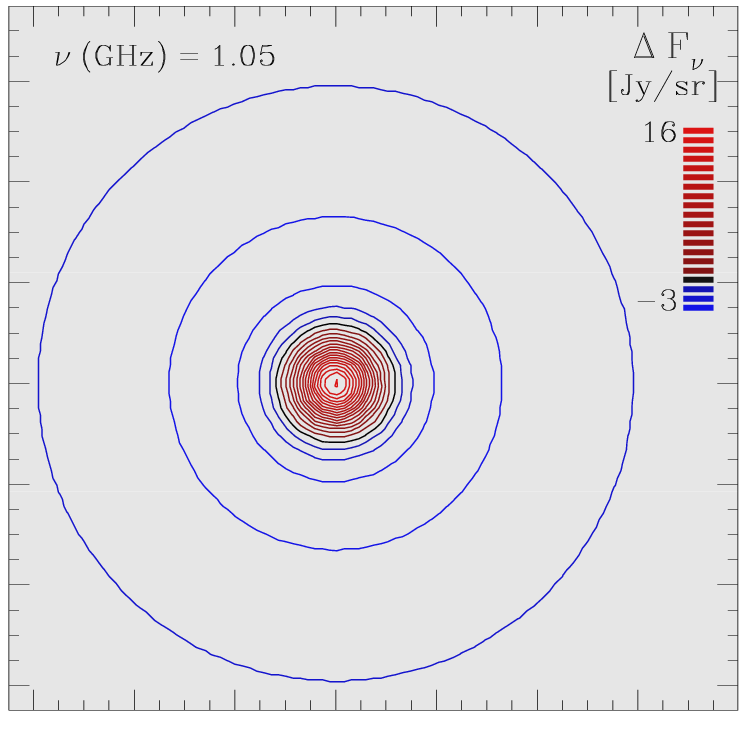}\\ [-1.7mm]
  \includegraphics[height=0.355028\linewidth]{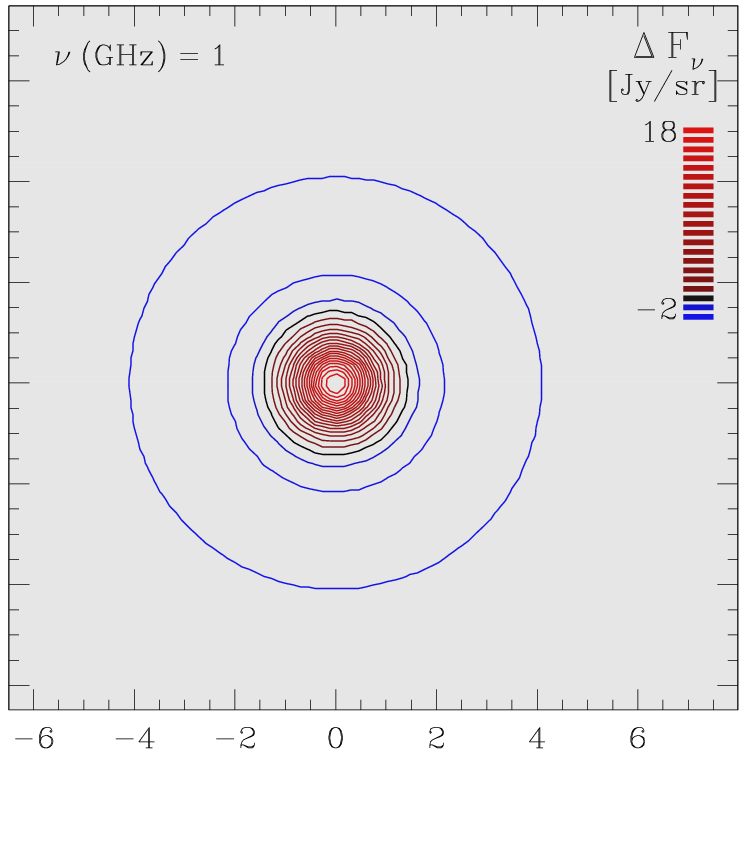}
  \hspace{-1.6mm}  
  \includegraphics[height=0.355028\linewidth]{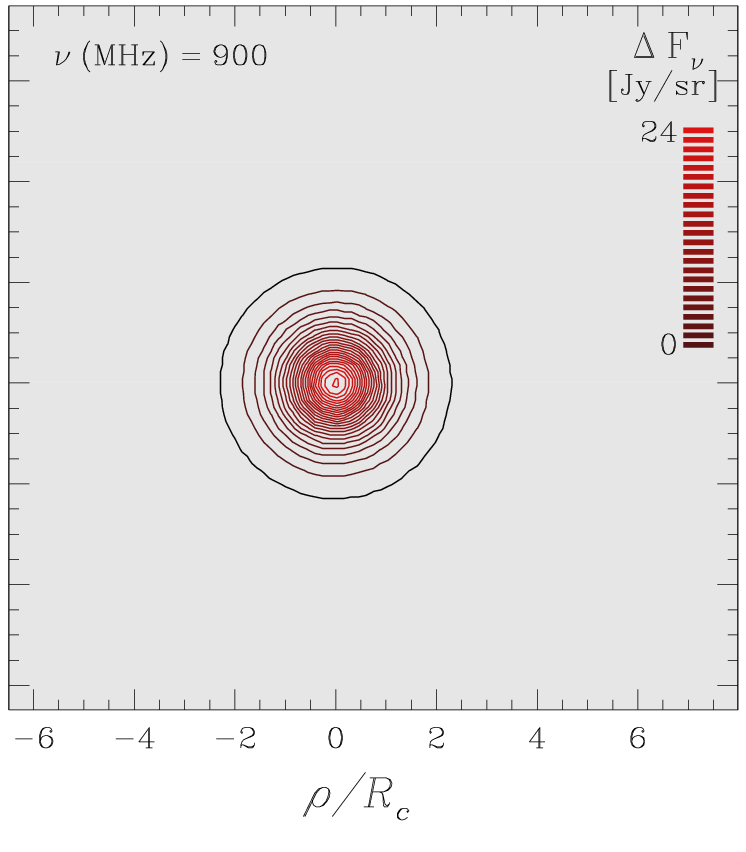}
  \hspace{-1.6mm}
  \includegraphics[height=0.355028\linewidth]{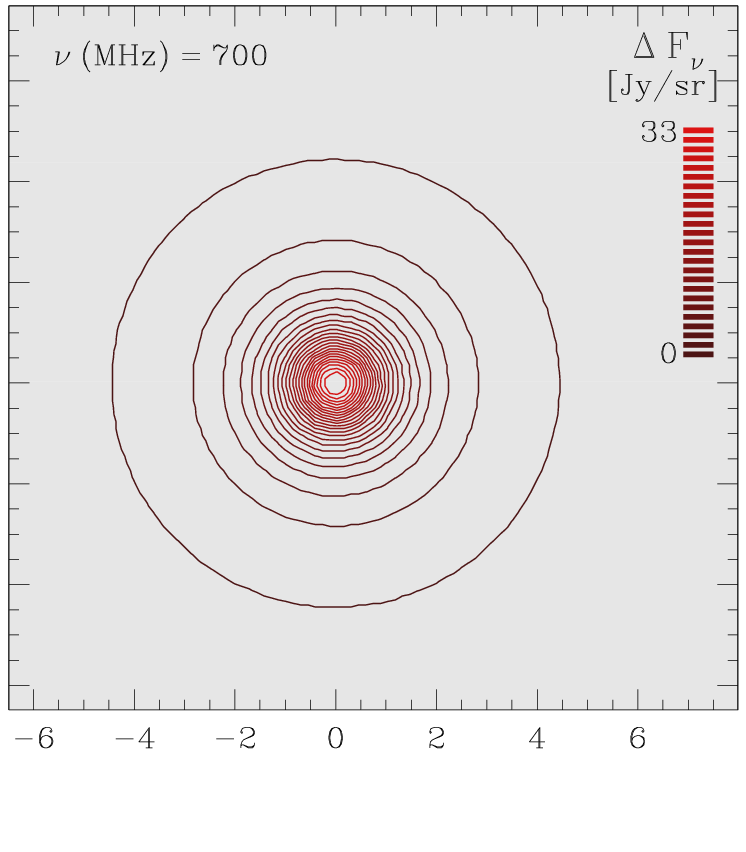}
  \end{minipage}\\
  
  \caption{\rm Evolution of the map of radio and microwave
    background distortions toward a galaxy cluster during the
    transition between the regimes of decreased and increased
    brightness (the blue and red lines indicate the ``negative''
    and ``positive'' background deviations, respectively). The
    distortions are caused by the scattering by electrons, but
    the appearance of a hybrid source (a bright spot surrounded
    by a dark ring) and a compact ``positive'' source on the map
    for $\nu=900$ MHz is largely associated with the thermal
    radiation of the hot cluster gas. The outer ($\rho \ga
    6\,R_{\rm c}$) contours on the maps with $\nu\leq1.0$ GHz
    are not shown. The computation for a nearby ($z\ll 1$)
    cluster with a $\beta$ density distribution ($\beta=2/3$)
    and the same parameters as those for the cluster in
    Figs.\,\ref{fig:beta-flux} and
    \ref{fig:maps}.\label{fig:maps-radio}}
\end{figure*}

The change in the shape of the source on the maps of background
fluctuations for this cluster during the transition near
$\nu\sim 1.1$ GHz from the completely ``negative'' source to the
completely ``positive'' one is shown in more detail in
Fig.\,\ref{fig:maps-radio}. It can be seen that the transition
occurs slowly, the hybrid source is observed in a wide frequency
range, $1.0\ \mbox{GHz}\la \nu \la 1.25\ \mbox{GHz}$, well above
the frequency $\nu_2\simeq 802$ MHz of the mutual compensation
of the Compton CMB and radio background distortions. The thermal
radiation of the hot gas is also seen at frequencies
$1.4\ \mbox{GHz}\la \nu \la 1.8\ \mbox{GHz}$ --- it forms a
central narrow peak inside the hole produced by the inverse
Compton scattering of the CMB. The hole at these frequencies
simply turns out to be deeper, and the thermal radiation peak
cannot rise above its edge and manifest itself as a clear
``positive'' central source. In fact, at these frequencies a
hybrid source, but in a latent form, is also observed at the
location of the cluster. Note also the map for $\nu=900$
MHz. The source at the location of the cluster on this map is
largely associated with the thermal radiation of the
intergalactic gas and has a noticeably smaller width than on
other maps, where it is formed primarily due to the Compton
scattering of the background. At $\nu=900$ MHz the distortions
due to the scattering of the radio and cosmic microwave
backgrounds compensate each other almost completely.

Surprisingly, when passing through $\nu_0\simeq 217.5$ GHz, the
source at the location of the cluster in Fig.~\ref{fig:maps}
also takes a hybrid shape. Previously, it has always been
thought that the ``negative'' source at the location of the
cluster near this frequency simply disappears to subsequently
appear at higher frequencies as a completely ``positive''
one. The frequency $\nu_0 = 3.8300\,kT_{\rm m}\simeq 217.5065$
GHz is calculated from the condition for the right-hand side of
Eq.~(\ref{eq:szcmb}), which describes the change in the CMB
spectrum upon Compton scattering, to become zero. The hybrid
shape of the source is again associated with the intergalactic
gas bremsstrahlung concentrated to the cluster center that
becomes noticeable due to the attenuation of the Compton CMB
distortions near this frequency. Note that the intensity levels
on the maps in Fig.~\ref{fig:maps} for $\nu=100$ and $300$ GHz
(immediately below and above $\nu_0$) are given in kJy/sr,
whereas those on the map with $\nu=217.47$ GHz and the remaining
maps are given in Jy/sr.
%
The change in the shape of the source at the location of the
cluster on the maps of CMB fluctuations when passing through
$\nu_0$ is studied in detail in Grebenev and Sunyaev (2024).

\begin{table*}[!t]
\caption{Parameters of the individual clusters selected to
  assess the possibility of measuring the effect of radio
  background scattering by electrons of their hot intergalactic
  gas\label{table:clusters}}

\vspace{3mm}
\begin{center}
\small
\begin{tabular}{l|l|l|c|c|c|c|@{\,}c@{\,}|c}\hline
\multicolumn{2}{c|}{}  & & & & & & &\\ [-3mm]
\multicolumn{2}{c|}{Cluster name\aa}&\multicolumn{1}{c|}{$z$\bb}&$R_{\rm c},$\cc&
$\theta_{\rm c},$\cc&$kT_{\rm e},$& $N_{\rm c}$\dd&$M_{\rm g}$\ee&$\tau_{\rm T}$,\ff\\ \cline{1-2} 
& & & & & & & & \\ [-3mm]
main&alternative&&kpc&\arcmin&keV&&$10^{14}\ M_{\odot}$& $10^{-3}$\\ \hline
 & & & & & & & &   \\ [-3.5mm]
 \multicolumn{1}{c|}{1}&\multicolumn{1}{c|}{2}& \multicolumn{1}{c}{3}&
 4    & 5&  6    &  7     &8  &  9    \\  \hline 
  & & & & &  & & &   \\ [-3.6mm]
AT\,J0102-4915&El Cordo        &0.870&270 &0.75 &14.5&8.9 &2.2 &15.2\\
A\,426&Perseus                 &0.018&280 &12.8 &6.0 &4.6 &2.0 &8.3\\
ST\,J0615-5746&P\,G266.6-27.3\,&0.972&230 &0.63 &14.2&7.2 &1.12&10.5\\ 
1E\,0657-558&Bullet            &0.296&170 &0.73 &12.4&12.3&2.01&13.3\\
A\,1656&Coma                   &0.023&290 &10.5 &6.9 &2.9 &1.0 &5.6\\
Virgo             &            &0.004&310 &62.5 &2.4 &2.7 &1.5 &5.3\\
A\,1991&                       &0.059&60  &0.90 &2.3 &6.4 &0.1 &3.5\\
ST\,J2106-5844&                &1.132&200 &0.54 &9.4 &11.5&1.17&14.6\\
ST\,J2248-4431&AS\,1063        &0.348&370 &1.4  &11.5&2.9 &1.95&7.0\\
ST\,J2344-4243&Phoenix         &0.596&290 &0.88 &14.9&4.8 &1.48&8.8\\ \hline
\multicolumn{9}{l}{}\\ [-1mm]
\multicolumn{9}{l}{\aa\ A -- Abell, ST -- SPT-CL, AT -- ACT-CL, P -- PLCK.}\\
\multicolumn{9}{l}{\bb\ The cluster redshift.}\\
\multicolumn{9}{l}{\cc\ The core radius in the model of a
  $\beta$ density distribution and its angular size.}\\ 
\multicolumn{9}{l}{\dd\ The gas density at the cluster center,
  $10^{-3}\ \mbox{\rm cm}^{-3}$.}\\
\multicolumn{9}{l}{\ee\ The gas mass in the cluster.}\\
\multicolumn{9}{l}{\ff\  The Thomson optical depth along the line
  of sight passing through the cluster center.} 
\end{tabular}
\end{center}
\vspace{-2mm}
\end{table*}

The detection of a hybrid source toward a galaxy cluster in both
CMB and radio background can be interesting not only as a
demonstration (and confirmation) of an elegant physical
effect. It may turn out to be important for a more accurate
separation of the radio background distortions due to the
Compton scattering and the contribution from the radio
bremsstrahlung of the cluster gas. One can attempt to estimate
the radio bremsstrahlung flux from X-ray cluster observations,
but the accuracy of such estimates depends critically on the
reliability of the determination of the gas temperature $kT_{\rm
  e}$. Direct measurements of the thermal radio flux from the
intergalactic gas may turn out to be more promising and
reliable.

\begin{figure*}[t!]
\hspace{7mm}\begin{minipage}{0.50\textwidth}
  \includegraphics[width=1.0\linewidth]{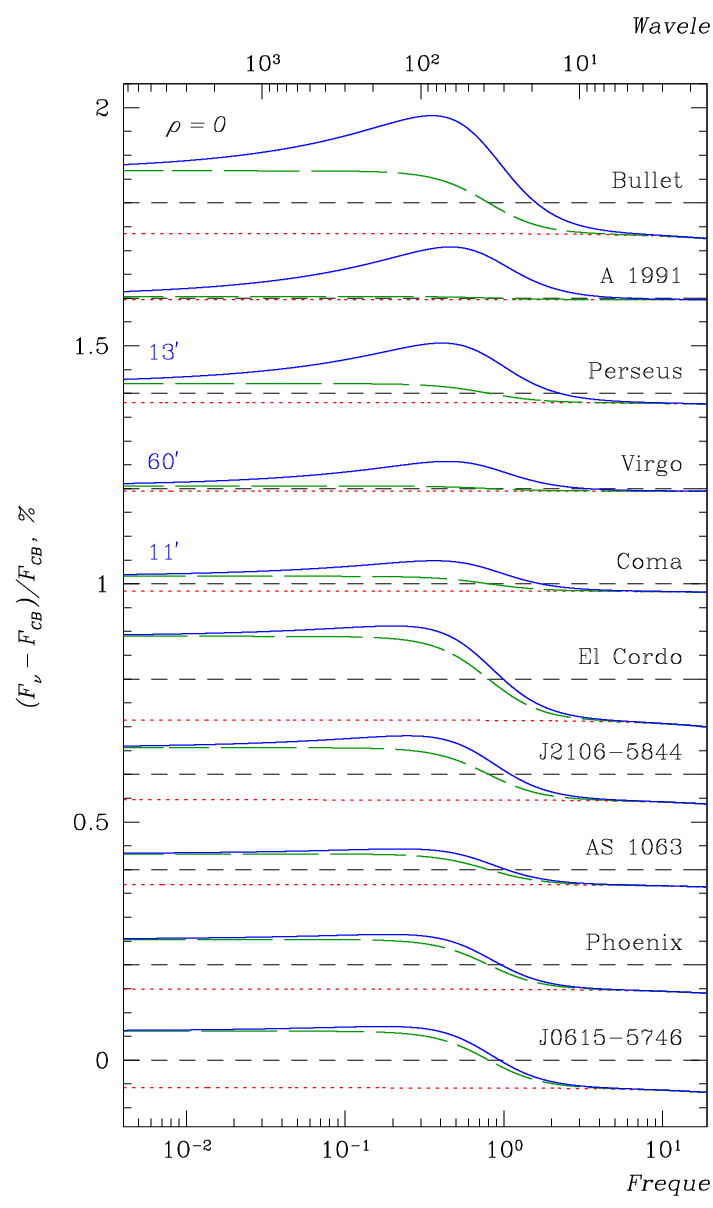}
\end{minipage}\hspace{-0.5mm}\begin{minipage}{0.50\textwidth}
  \includegraphics[width=1.0\linewidth]{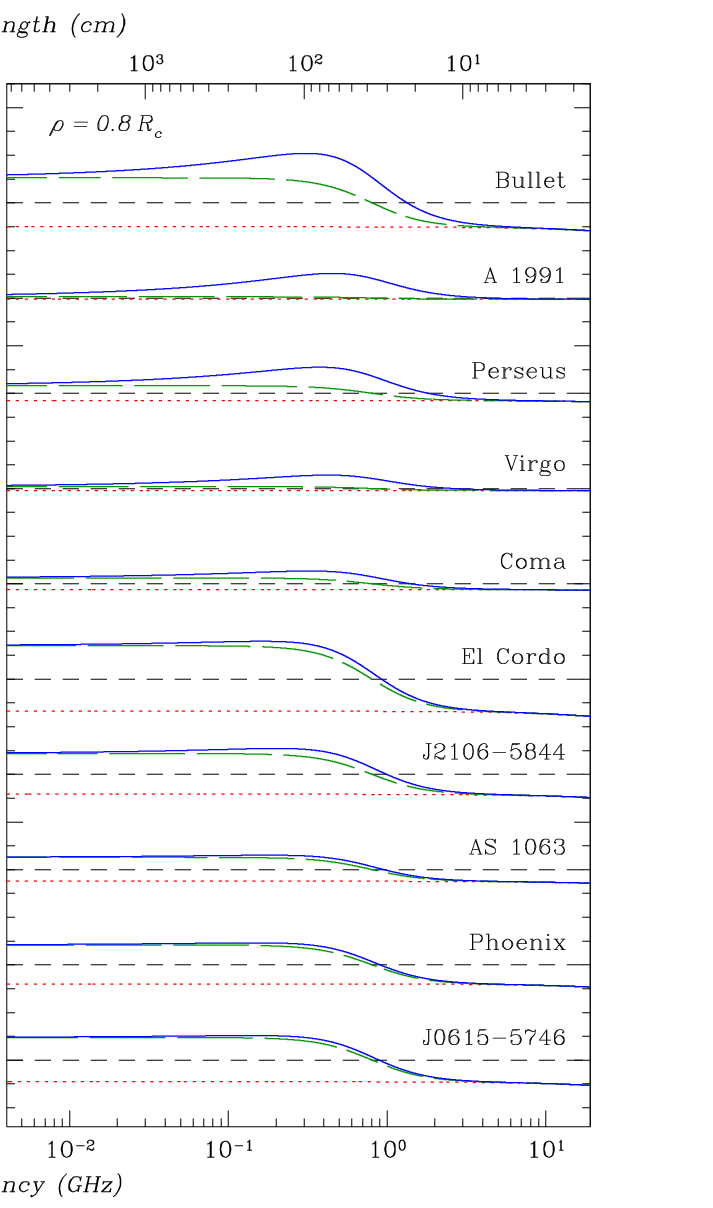}
\end{minipage}
\caption{\rm The relative (in \%) distortions in the spectrum of
  the cosmic radio and microwave background expected due to its
  scattering by hot-gas electrons in several known galaxy
  clusters (long green dashes). The blue lines also take into
  account the thermal (bremsstrahlung) radiation of the gas. The
  decrease in CMB brightness due to its scattering by electrons
  is indicated by the red dotted lines. The distortions observed
  toward the cluster center ({\sl left\/}) and at an impact
  parameter close to the cluster core radius $\rho = 0.8\,R_{\rm
    c}$ ({\sl right}). The gas is assumed to be isothermal and
  to have a $\beta$ density distribution ($\beta=2/3$), the
  cluster parameters are given in
  Table\,\ref{table:clusters}.\label{fig:clusters}}
\end{figure*}

\subsubsection*{\uwave{Individual clusters.}\/}~Let us assess the possibility
of detecting the distortions of the radio background associated
with its scattering by intergalactic gas electrons toward
several known galaxy clusters. We will assume the gas to be
isothermal and to have a $\beta$ density distribution with
$\beta=2/3.$ We will make our estimates for rich clusters that
manifest themselves by strong microwave background distortions
from the sample used in Grebenev and Sunyaev
(2019). Table\,\ref{table:clusters} gives the main
characteristics of the intergalactic gas (its temperature and
central density, other parameters of the $\beta$-model) and the
clusters themselves (redshift $z$).

The long green dashes in Fig.\,\ref{fig:clusters} indicate the
expected relative (in \%) Compton distortions in the radio and
microwave background spectrum for several clusters from this
sample. The observations are assumed to be carried out toward
the cluster center (the left panel of
Fig.~\ref{fig:clusters}) or at an impact parameter
$\rho=0.8\ R_{\rm c}$ relative to the center (its right
panel). The unperturbed background level is marked by the short
black dashes.  The red dotted lines indicate the contribution
from the effect of a decrease in brightness due to the
scattering of only the CMB to the distortion spectrum. It can be
seen that the relative Compton distortions do not depend on the
frequency in the case of both the radio background and the CMB
(in complete agreement with the predictions of
Eqs.\,[\ref{eq:szcmb}] and [\ref{eq:szcrb}]). Moreover, they are
almost equal in magnitude, but have different signs. The blue
solid lines indicate the background distortions including the
intrinsic bremsstrahlung of the hot cluster gas.

Figure~\ref{fig:clusters} shows that the scattering of the
radio emission in a number of nearby clusters leads either to
very small (the Coma and Perseus clusters) or negligible (the
Virgo and A\,1991 clusters) background distortions. The low
temperatures ($kT_{\rm e}\sim 2.5-6.9$ keV) and quite moderate
optical depths ($\tau_{\rm T}\la 8\times 10^{-3}$) of the gas
contained in them have an effect. At the same time, the radio
bremsstrahlung of these nearby ($z\la 0.06$) clusters is fairly
strong. This manifests itself particularly clearly on the left
panel of the figure, in the observations toward the cluster
center. The detectability of the radio background distortions in
the first two clusters can be increased by integrating the radio
signal over their fairly large apparent area (the angular size
of these clusters is specified in Fig.~\ref{fig:clusters} on
the left). As has already been mentioned (and will be shown
below), the contribution of the scattered radiation to the
overall spectrum of the radio background distortions increases
compared to the contribution of the bremsstrahlung (due to the
strong concentration of the latter to the cluster center).
Another possibility is to observe the distortions at an
appreciable impact parameter $\rho \sim R_{\rm c}$ from the
cluster center, as shown on the right panel of
Fig.~\ref{fig:clusters}. It can be seen that even at an
impact parameter $\rho = 0.8\ R_{\rm c}$ the bremsstrahlung
intensity is a factor of $\sim 2.1$ lower than the intensity
toward the cluster center, with the amplitude of the Compton
background distortions changing not too dramatically ($<30$\%).
 
Nevertheless, it is obvious that to detect the Compton
distortions, it is much more promising to observe very hot
($kT_{\rm e}\ga 12$ keV), optically thick ($\tau_{\rm T}\ga
8\times 10^{-3}$), distant ($z\ga 0.35$) clusters: El Cordo,
Phoenix, AS~1063, ST~J0615-5746 and ST~J2106-5844. The Compton
background distortions in such clusters can reach $\sim0.1$\%;
the main obstacle for their measurement in the form of hot-gas
bremsstrahlung weakens noticeably due to the great distance of
the clusters (high $z$). The Bullet cluster, one of the hottest
and optically thickest clusters, but, at the same time, with
powerful thermal gas radiation, is an exception.  This is partly
because of its moderate redshift $z\simeq 0.3$, and partly
because of its unusual compactness, $R_{\rm c}\simeq 170$ kpc,
and, as a consequence, the enhanced density of its gas $N_{\rm
  c}\simeq 1.23\times 10^{-2}\ \mbox{\rm cm}^{-3}$. Even in
the decameter wavelength range and in the case of
peripheral observations, it will be difficult to detect the
Compton background distortions toward this cluster.

\subsubsection*{\uwave{Observations of unresolved
    clusters.}\/}~The angular size of many distant clusters is
too small to perform their peripheral observations. In this
case, their entire radio flux or the integrated flux from the
wide central part of the cluster is measured.
\begin{figure*}[!t]
  \hspace{3mm}\includegraphics[height=0.77\linewidth]{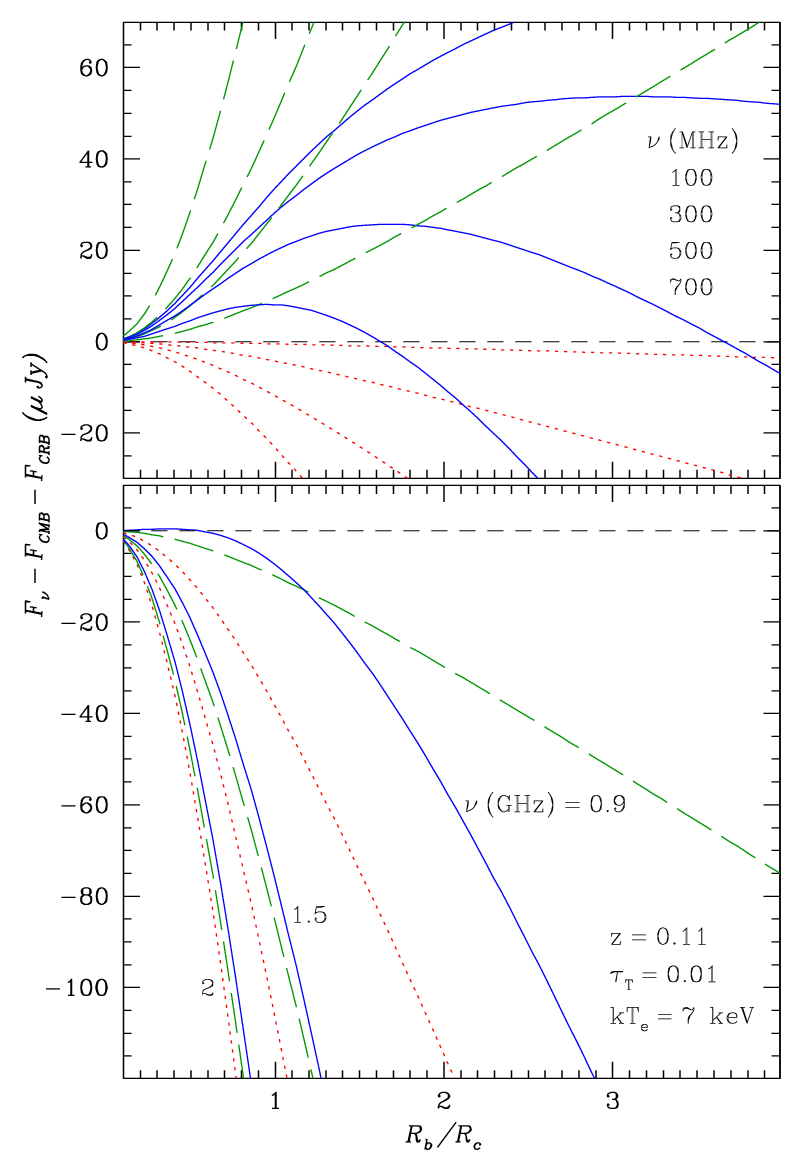}
  \hspace{-3mm}
  \includegraphics[height=0.77\linewidth]{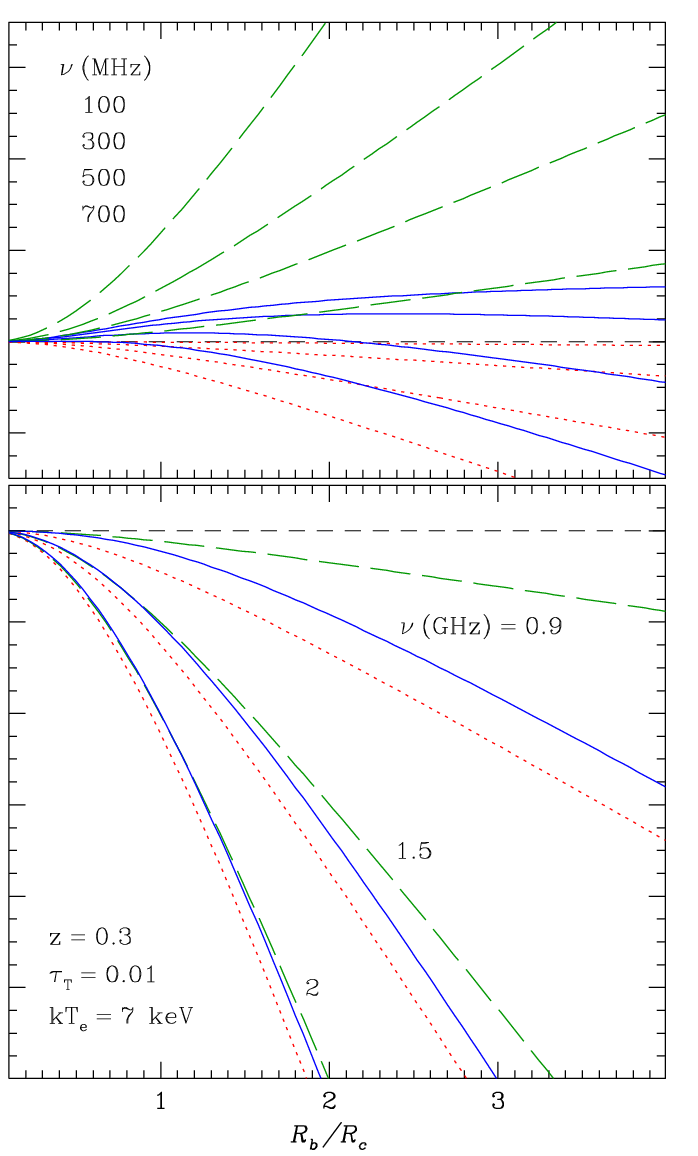}
\caption{\rm The integrated radio flux within the radius $R_{\rm
    b}$ from a galaxy cluster with a $\beta$ intergalactic gas
  density distribution ($\beta = 2/3$), a core radius $R_{\rm c}
  = 350$ kpc, and a Thomson optical depth toward the center
  $\tau_{\rm T} = 0.01$ (see Figs.~\ref{fig:dev2}a and
  \ref{fig:beta-flux}). Such a flux will be recorded by a
  telescope with an angular resolution $\theta_{\rm b}=R_{\rm
    b}/d_{\rm A},$ where $d_{\rm A}$ is the angular-diameter
  distance to the cluster. We assume the gas to be isothermal
  with a temperature $kT_{\rm e} = 7$ keV, and the cluster
  itself to be at $z=0.11$ ({\sl left}) or $0.3$ ({\sl
    right}). The radio emission is associated with (1) the
  scattering of the cosmic radio background by electrons of the
  hot gas (long green dashes) or (2) the thermal bremsstrahlung
  of this gas (blue solid lines). In both cases, we took into
  account the CMB distortions due to the Compton scattering (red
  dotted lines). The lines of the same type correspond to
  different radio frequencies.\label{fig:kingtot}}
\end{figure*}

Take a cluster with a $\beta$ hot-gas density distribution and a
core radius $R_{\rm c}= 350$ kpc. Consider two cases where the
cluster is located at $z=0.11$ or $0.3$. The corresponding
angular-diameter distances\footnote{We adopted the standard
  $\Lambda$CDM cosmological model with $\Omega_{\rm M} = 0.3$,
  $\Omega_{\rm \Lambda} = 0.7,$ and $H = 70\ \mbox{km s}^{-1}
  \mbox{\rm Mpc}^{-1}.$} to the cluster are $d_{\rm A}\simeq
390$ or $810$ Mpc; the angular size (radius) of its core is
$\theta_{\rm c}\simeq 3\farcm0$ or $1\farcm5$. Let us integrate
Eqs.~(\ref{eq:kingme}) and (\ref{eq:kingtau}) over the cluster
area $2\pi \int^{R_{\rm b}}_0 \rho\,d\rho/d_{\rm A}^2$ within
the solid angle limited by the half-angle $\theta_{\rm b}=R_{\rm
  b}/d_{\rm A}$ to find the integrated bremsstrahlung and
Compton parameters of the cluster:
$$  Y_{\rm B}(R_{\rm b})=1.18\,\pi^2 N_{\rm c}^2 R_{\rm
    c}\left(\frac{R_{\rm c}}{d_{\rm A}}\right)^2  
  \left[ 1-\left(1+\frac{R_{\rm b}^2}{R_{\rm c}^2}\right)^{-1/2}
    \right]
$$
and
$$ Y_{\rm C}(R_{\rm b})=2\pi\tau_{\rm T} \left(\frac{kT_{\rm e}}{m_{\rm
   e}c^2}\right) \left(\frac{R_{\rm c}}{d_{\rm A}}\right)^2
 \left[\left(1+\frac{R_{\rm b}^2}{R_{\rm c}^2}\right)^{1/2}\!-1\right].\nonumber
$$
In particular, it can be seen that even at $R_{\rm
  b}=0.8\ R_{\rm c}$ the ratio of the integrated Compton
parameter $Y_{\rm C}\simeq 0.44\ y_{\rm C}(0) (\pi \theta_{\rm
  c}^2)$ to the integrated bremsstrahlung parameter of the
cluster $Y_{\rm B}\simeq 0.21\ y_{\rm B}(0) (\pi \theta_{\rm
  c}^2)$ is a factor of 2.1 greater than the ratio of their
local values $y_{\rm C}(0)/y_{\rm B}(0)$ toward the cluster
center. The bremsstrahlung, $y_{\rm B}(0)$, and Compton, $y_{\rm
  C}(0)$, parameters toward the cluster center can be found from
Eqs.~(\ref{eq:kingme}) and (\ref{eq:kingtau}).

The actual ratio of the integrated fluxes for the radio emission
that is associated with the Compton scattering of the background
or is the intrinsic bremsstrahlung of the hot gas depends on the
radio frequency at which the measurements are carried out.
Figure~\ref{fig:kingtot} shows how the integrated radio flux from
the galaxy cluster under consideration is gained at various
frequencies as $R_{\rm b}$ increases (as the angular resolution
of the telescope used for the observations degrades). The
telescope is assumed to be pointed to the cluster center; the
half-width of its point spread function is $\theta_{\rm
  b}=R_{\rm b}/d_{\rm A}$. The left and right panels in the
figure show the cases where the cluster is located at $z=0.11$
and $z=0.3$, respectively; the other cluster parameters and the
intergalactic gas parameters were taken to be the same as those
in Figs.~\ref{fig:dev2}a and \ref{fig:beta-flux}.

Irrespective of the nature of the radiation, the pre- sented
fluxes take into account the scattering of CMB photons in the hot
intergalactic gas (its contribution is indicated by the red
dotted line). The drop in the recorded bremsstrahlung flux (and
at high frequencies also the scattered radio background flux) as
the angular resolution of the telescope degrades is associated
precisely with the contribution of the CMB deficit. At low
frequencies the scattered flux, on the contrary, increases with
degrading resolution, particularly dramatically when observing
a more nearby cluster. The point is that (1) the contribution of
the CMB at these frequencies drops rapidly and (2) the figure
presents the flux integrated over the cluster area, while the
integration angle increases with increasing ratio
$R_{\rm b}/R_{\rm c}$ and decreasing $z$.

It follows from the figure that in nearby clusters (at $z\la
0.11$) at low frequencies $\nu\la 500$ MHz the contribution of
the Compton scattering of the radio background dominates over
the thermal gas radiation irrespective of the telescope's
resolution. At frequencies $500\ \mbox{\rm MHz}\la\nu\la 1$ GHz
the Compton scattering prevails only if the point spread
function of the telescope covers an extended region of the
cluster ($R_{\rm b}\ga 1.5-2\ R_{\rm c}$). The reason is that
the bremsstrahlung flux at these frequencies, after a slight
initial growth, then decreases with increasing $R_{\rm b}$,
whereas the contribution of the scattering changes monotonically
with increasing $R_{\rm b}$. At frequencies $\nu\ga 1$ GHz the
thermal gas radiation makes a major contribution to the measured
radio flux (an excess relative to the decrease in CMB brightness
associated with its scattering), since the radio background
weakens.

At the same time, in distant clusters ($z\ga0.3$) the Compton
scattering of the radio background dominates over the thermal
gas radiation at all of the frequencies considered up to
$\nu\sim2$ GHz.
\begin{figure*}[!t]
\centering
\includegraphics[width=0.85\linewidth]{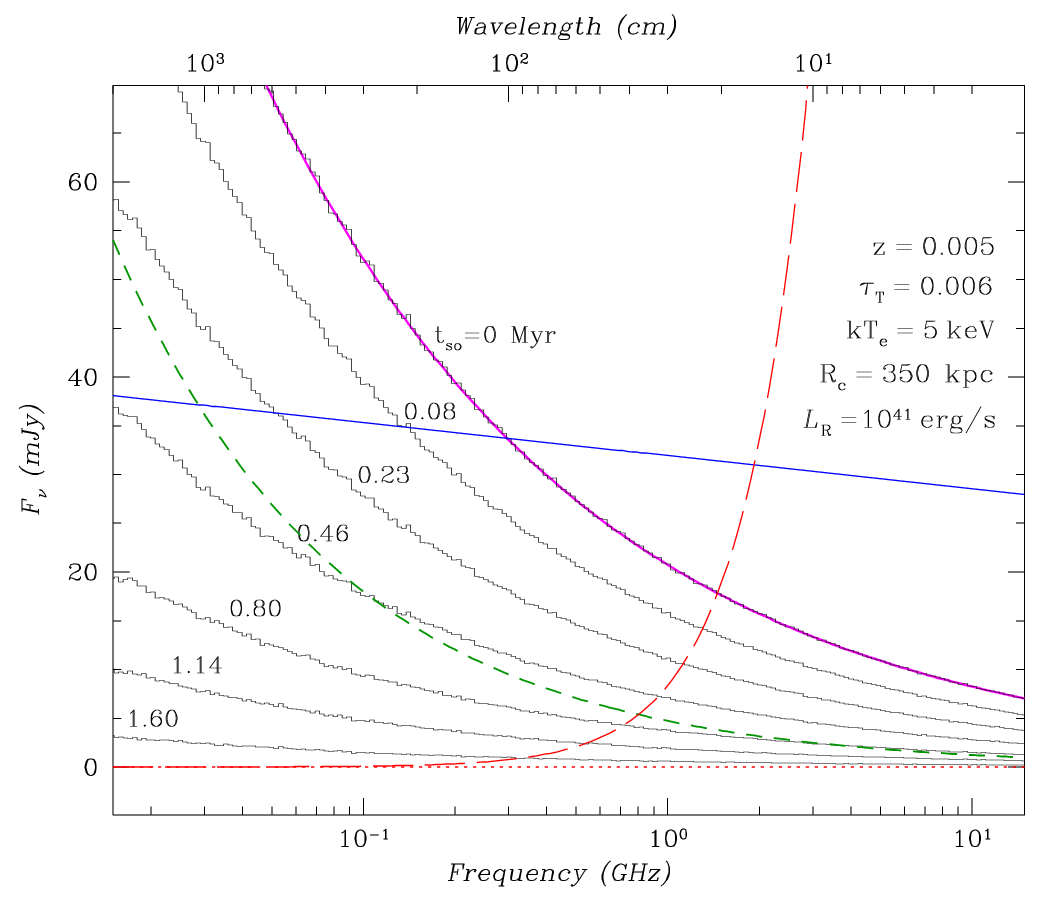}
\caption{\rm The distortions in the spectrum of the cosmic radio
  background due to its scattering by electrons of the hot
  cluster gas (green dashes), the bremsstrahlung of this gas
  (blue solid line), and the scattered (diffuse) radiation of
  the central galaxy (histograms). The red curve (long dashes)
  indicates the absolute value of the decrease in CMB brightness
  due to its scattering in the cluster gas. The integrated flux
  (from the entire cluster) is presented everywhere. It is
  assumed that the galaxy had been active for a long time in the
  radio band with a spectral index $\gamma=0.4$ and a luminosity
  $L_{\rm R}=1\times 10^{41}\ \mbox{erg s}^{-1}$ in the
  frequency range 10 MHz -- 100 GHz, but switched off $t_{\rm
    so}$ million years ago. The hystograms are obtained from
  Monte Carlo computations. The case of $t_{\rm so}=0$ was also
  calculated analytically (Eq.~[\ref{eq:agn}], the dark-red
  line). We consider a nearby ($z=0.005$) cluster with a uniform
  isothermal ($kT_{\rm e}=5$ keV) gas density, a radius $R_{\rm
    c}=350$ kpc, and an optical depth along the line of sight
  $\tau_{\rm T}=0.006$ (the same as that in
  Fig.~\ref{fig:dev1}).\label{fig:agn}}
\end{figure*}

\section*{RADIO EMISSION FROM CLUSTER GALAXIES}
\noindent
Cluster galaxies often possess a powerful radio emission that
can have a complex morphology (see, e.g., Hill and Longair 1971;
Cuciti et al. 2018) associated with the interaction of the
relativistic jets outflowing from the galaxies and expanding
plasma bubbles with the surrounding intergalactic gas, the
acceleration and ejection of relativistic particles into the
surrounding medium, and the generation of synchrotron radiation
by them. Irregularities in the radio emission distribution are
most typical for interacting, colliding galaxy clusters; they
primarily manifest themselves near the shocks and contact
discontinuities arising in this case.

The nonuniformity of the distribution of the synchrotron radio
emission from clusters can be used for the estimation of its
intensity and the proper subtraction. Similarly, if the cluster
galaxies continue to remain active in the radio band at the time
of their observations, then their contribution to the increase
in the radio background flux observed toward the cluster can be
estimated and taken into account.

\subsubsection*{\uwave{An echo of the past activity of radio galaxies.}\/}~The situation is different if one or more cluster
galaxies were active in the past and are now in the ``off'' or
``radio-quiet'' states. Their radio emission at the epoch of
activity scattered in the hot cluster gas can be observed as
diffuse cluster radiation even now, leading to an additional
increase in its brightness relative to the background level. The
point is that the scattered emission traverses a much longer
path inside the cluster than does the direct radiation from the
galaxies and comes to us with a noticeable time delay due to the
large cluster sizes (hundreds of parsecs). Besides, it is
distributed more uniformly over the intergalactic gas cloud than
is the direct radiation from the galaxies. Having a synchrotron
nature, this emission is described by a power-law spectrum close
to the radio background spectrum. As a result of all these
factors, the scattered emission of the past activity of galaxies
can make it much more difficult to detect and identify the
excess of the radio background associated with its scattering in
the hot cluster gas.

In contrast to the Compton scattering of the isotropic
background radiation whose distortions arise due to the
relativistic (Doppler) effects $\sim kT_{\rm e}/mc_{\rm e}^2$,
the diffuse radiation from active galaxies is determined mainly
by the change in the direction of photon motion upon Thomson
scattering. The change in the photon frequency upon scattering
affects the spectrum of the diffuse radiation insignificantly
(at the same level $\sim kT_{\rm e}/mc_{\rm e}^2\sim 1$--2\%,
see Eq.~[\ref{eq:szcrb}]).

If an active galaxy is at the cluster center and switched off
only recently, $t_{\rm so}\ll R_{\rm c}/c$, while before this it
shined for a long time in the radio band at approximately the
same level and had a power-law spectrum with a spectral index
$\gamma$, then the integrated, direction-averaged, spectral
density of the scattered emission from the cluster can be easily
calculated:
\begin{equation}\label{eq:agn}
  F_{\nu}(\nu)=F_{\rm A}(\nu)
  \tau_{\rm c}\, R_{\rm c}^2/d_{L}^2,\  \mbox{\rm where}
\end{equation}
\begin{equation}\label{eq:agn0}
F_{\rm A}(\nu)=\frac{L_{\rm R}}{4\pi R_{\rm c}^2}\
\left(\frac{(1-\gamma)\ 10^{14}}{10^{(2-2\gamma)}-10^{(2\gamma-2)}}\right)
\nu_9^{-\gamma}\ \mbox{Jy}.
\end{equation}
Here, $\nu_9$ is the frequency in GHz, $L_{\rm R}$ is the radio
luminosity of the galaxy in erg s$^{-1}$ in the range 10 MHz --
100 GHz before the switch-off, and $d_{L}(z)$ is the luminosity
distance at which it is located.

Figure~\ref{fig:agn} presents the results of such calculations
performed under the assumption that $\gamma=0.4,$ $L_{\rm R} =
1\times 10^{41}\ \mbox{\rm erg s}^{-1}$, and the galaxy cluster
is at $z=0.005,$ i.e., at a luminosity distance $d_{\rm
  L}=21.44$ Mpc (an angular-diameter distance $d_{\rm A}=21.23$
Mpc). The cluster radius $R_{\rm c}=350$ kpc corresponds to an
angular size $\theta_{\rm c}=57\arcmin$; the electron density
distribution is assumed to be uniform. The remaining cluster
parameters were taken to be the same as those in
Fig.\,\ref{fig:dev1}. The model cluster with such parameters
resembles the Virgo cluster, but contains a hotter gas.

The result of our calculation using Eq.~(\ref{eq:agn}) is
indicated by the dark-red line. The histograms represent the
results of analogous Monte Carlo computations using the
algorithms of Pozdnyakov et al. (1983). We used the computer
code developed by us when working on our previous paper
(Grebenev and Sunyaev 2019). The analytical curve is seen to
agree excellently with our numerical computation for the galaxy
switch-off at $t_{\rm so}=0$. Other histograms indicate the
spectrum of the diffuse radiation that must be recorded from the
cluster a certain time, $t_{\rm so}=0.08,$ 0.23, 0.46, 0.80,
1.14, and 1.60 Myr, after its switch-off. Note that the last
photon of the intrinsic (direct) radiation from the galaxy
leaves the hot-gas cloud in the cluster $R_{\rm c}/c\simeq 1.14$
Myr after its switch-off. We must record the spectrum at
precisely this time. The scattered photons of the last two
histograms spent more than twice as long in the cloud.
\begin{figure*}[!t]
\hspace{-4.5mm}\begin{minipage}{1.03\linewidth}
  \includegraphics[height=0.298\linewidth]{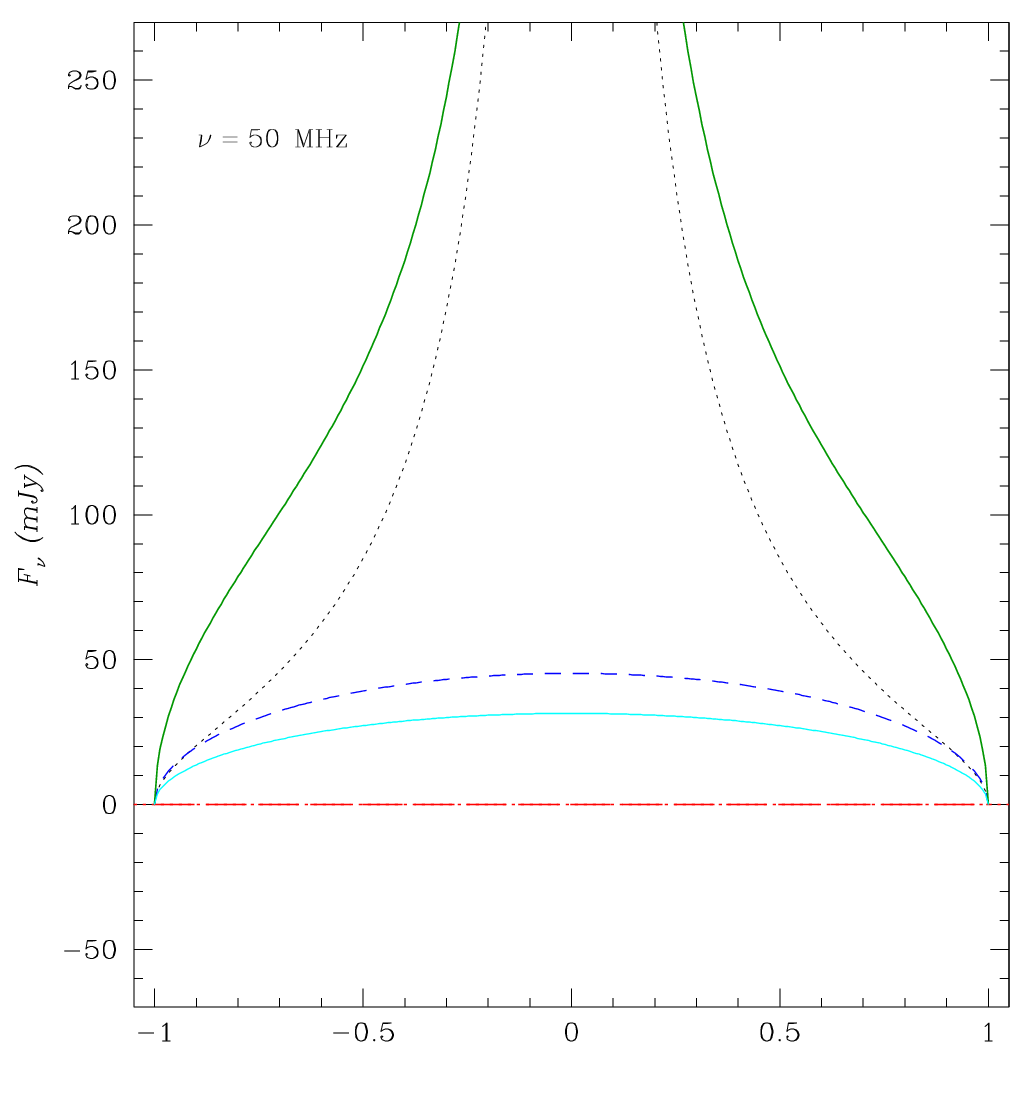}\hspace{-1mm}
  \includegraphics[height=0.298\linewidth]{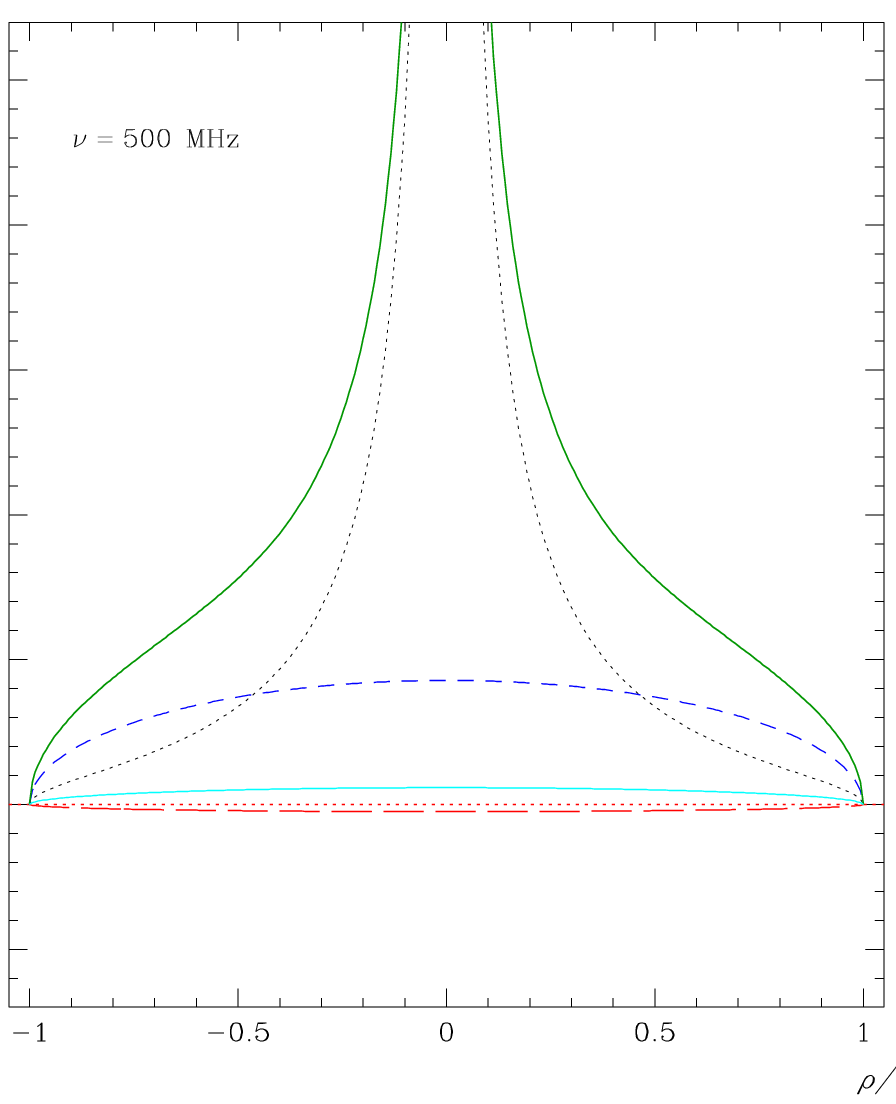}\hspace{-1mm}
  \includegraphics[height=0.298\linewidth]{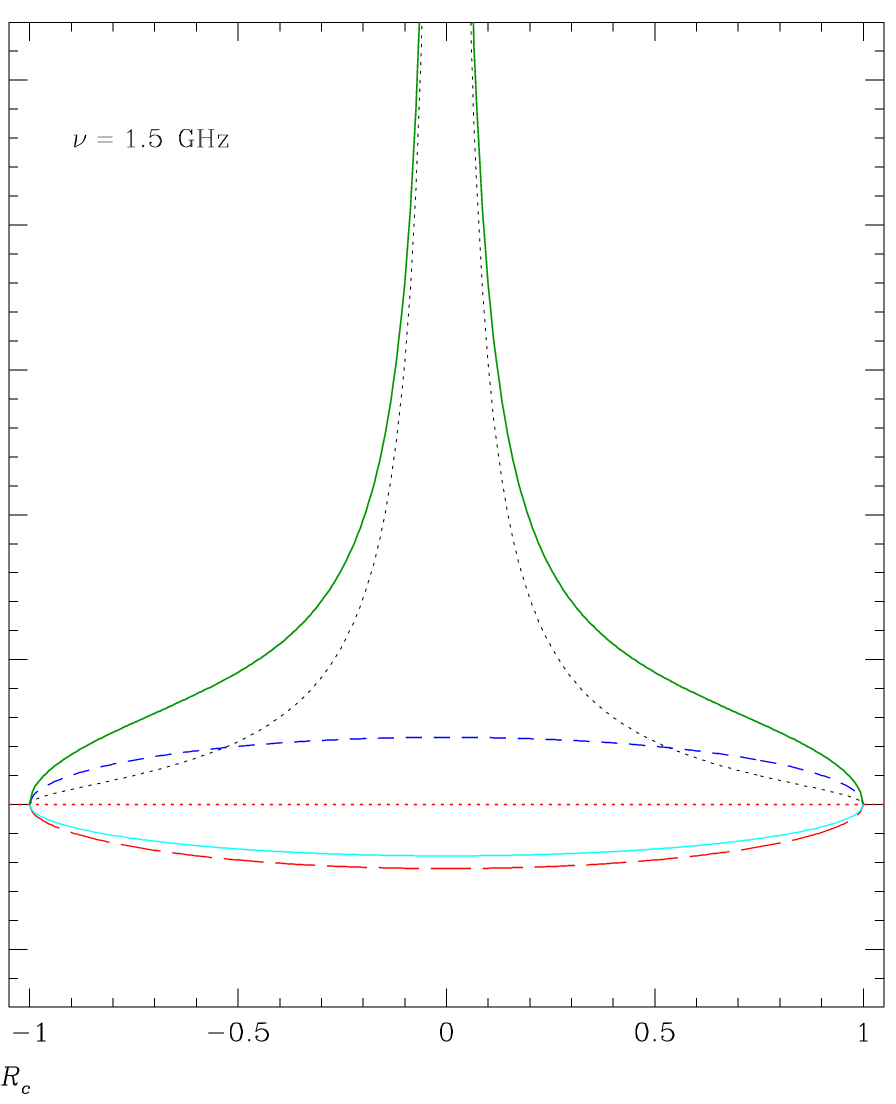}\hspace{-1mm}
  \includegraphics[height=0.298\linewidth]{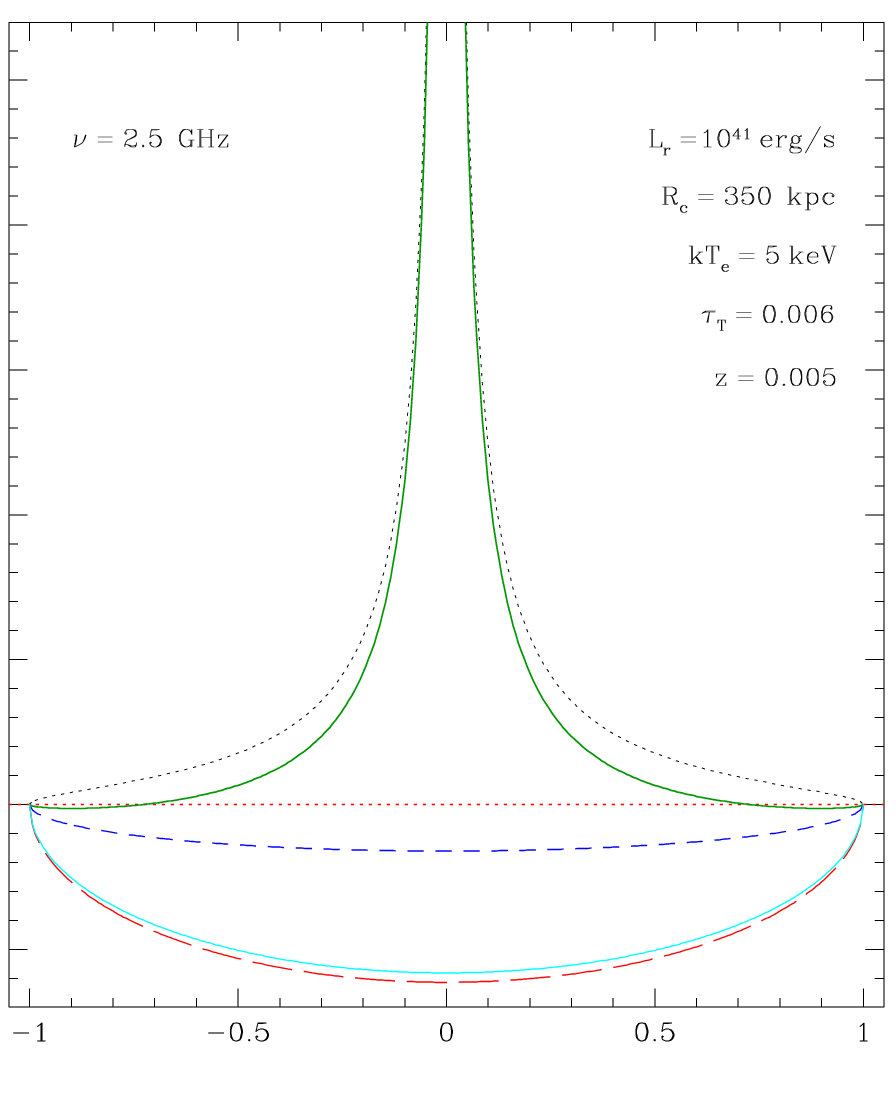}
  \end{minipage}
\caption{\rm Cosmic radio background excess toward a galaxy
  cluster versus impact parameter $\rho$ (green solid line). The
  cyan solid line indicates the distortions due to the
  scattering in the hot cluster gas, the short blue dashes
  indicate the contribution of the gas bremsstrahlung, and the
  black dotted line indicates the scattered emission from the
  central galaxy. Everywhere, except for the galaxy emission, we
  took into account the decrease in the brightness of the CMB
  due to its Compton scattering (long red dashes). It is assumed
  that the galaxy was active for a long time and had a spectral
  index $\gamma=0.4$ and a luminosity $L_{\rm R}=1\times
  10^{41}\ \mbox{erg s}^{-1}$ in the range 10 MHz -- 100 GHz, but
  recently ($t_{\rm so}\simeq0$) it switched off. The cluster is
  considered with the same parameters as those in
  Fig.~\ref{fig:agn}.\label{fig:agnbeam}}
\end{figure*}

Figure~\ref{fig:agn} also presents the spectral density of the
thermal (bremsstrahlung) radiation from the hot ($kT_{\rm e}=5$
keV) cluster gas (blue solid line) and the excess radiation
associated with the Compton scattering of the radio background
(green dashes). In contrast to most of the previous figures, the
fluxes from the entire cluster obtained by the integration over
its apparent solid angle $2\pi\int_0^{R_{\rm c}}\rho\,
d\rho/d_{\rm A}^2$ are presented here. Also, in contrast to other
figures, here we ignored the decrease in CMB
brightness. Instead, the absolute value of this decrease is
shown (long red dashes). Clearly, at frequencies above
$\sim1-1.5$ GHz all of the excess radiation associated with the
past activity of the galaxy, along with the radiation associated
with the radio background scattering and the cluster gas
bremsstrahlung, are completely suppressed by the decrease in CMB
brightness. At the same time, at frequencies $\nu\la 300$ MHz
the diffuse radiation from the recently ($\la 0.5$ Myr ago)
switched-off galaxy with the above luminosity can dominate in
the radio background excess being detected toward the cluster,
exceeding both the background scattering effect and the
contribution of the thermal gas radiation. In the frequency
range $300\ \mbox{\rm MHz} \la\nu\la 1.5\ \mbox{\rm GHz}$ the
thermal intergalactic gas radiation dominates in the cluster
with the chosen parameters.

Figure~\ref{fig:agn} shows that the scattered emission from the
cluster galaxy that was active in the past may turn out to be a
serious obstacle in the way of detecting (to be more precise,
identifying) the excess of the radio background associated with
its Compton scattering by electrons of the hot intergalactic gas
in this direction. Of course, the case considered suggests that
the galaxy was very bright in the radio range in the recent
past. However, for a massive central cluster galaxy such
activity cannot be ruled out.

The fact that the the diffuse radiation in distant clusters at
high redshifts $z$ must be greatly attenuated compared to the
scattered background radiation (independent of $z$) causes some
optimism. Above we showed how rapidly the intensity of the
thermal radiation from the hot cluster gas decreases with
increasing $z$. The problem is that the spectra of the scattered
background radiation and the diffuse radiation from an active
galaxy are very close in shape (dependence on $\nu$) and they
are difficult to distinguish.

\subsubsection*{\uwave{Morphology of the diffuse radiation from a
radio galaxy.}\/}~Certain hopes for revealing the scattered
background radiation can be associated with different
morphologies of the distribution of its brightness and the
brightness of the diffuse radiation from a switched-off galaxy
in the picture plane of the cluster. Indeed, when observing a
cluster with a uniform hot-gas density distribution using an
antenna with a good angular resolution 
($\ll R_{\rm c}/d_{\rm A},$ where $d_{\rm A}$ is the
angular-diameter distance to the cluster), the background excess
flux, along with the intensity of the thermal gas radiation,
depend on the impact parameter $\rho$ proportionally to
$(1-\rho^2/R_{\rm c}^2)^{1/2}$. The situation is different for
the scattered emission from a radio galaxy that switched off
only recently. 
\begin{figure*}[!t]
\centering
\includegraphics[width=0.82\linewidth]{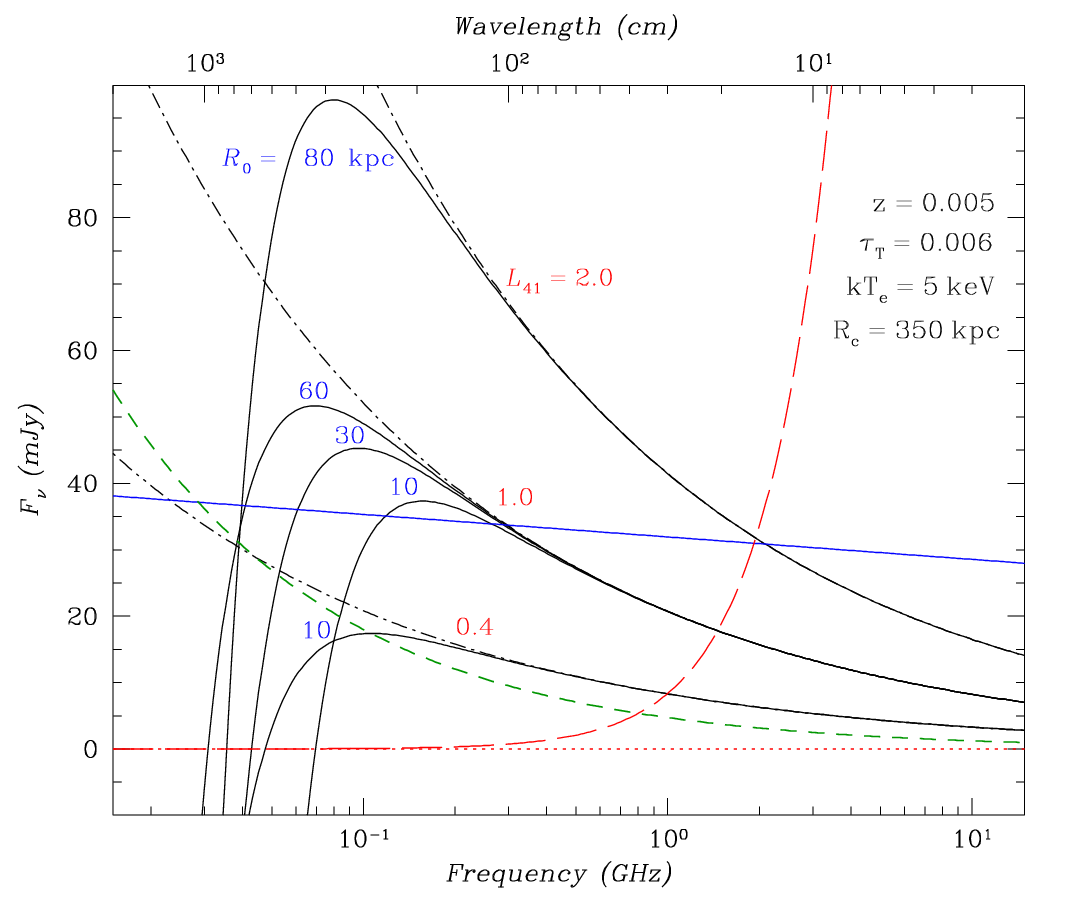}
\caption{\rm Same as Fig.~\ref{fig:agn}, but for central
  galaxies of various luminosities in the frequency range 10 MHz
  -- 100 GHz, $L_{\rm R}=L_{41}\times 10^{41}\ \mbox{erg
    s}^{-1}$, where $L_{41}=2,\ 1,\ \mbox{or}\ 0.4$ (indicated by
  the red numbers), that switched off only recently, $t_{\rm
    so}=0$. The radio emission from the galaxies scattered in
  the gas is indicated by the black dash-dotted lines. The black
  solid lines indicate the scattered emission from the galaxies
  including the induced Compton scattering. It is assumed that
  the scattering gas is absent within the radius $R_0=10-80$ kpc (blue
  numbers) near the galaxies.\label{fig:agnind}}
\end{figure*}

Although the integrated scattered flux is gained with radius (to
be more precise, with radial optical depth, see
Eq.~[\ref{eq:agn}]) linearly (the fraction of the emission
scattered in a spherical layer at a given radius is determined
only by its optical depth $\Delta\,\tau_{\rm T}$),
the intensity of the scattered emission obtained by the
integration along the line of sight at an impact parameter
$\rho$ depends on $\rho$ in a more complicated way:
\begin{equation}\label{eq:agnbeam}
  F_{\nu}(\nu)= F_{\rm A}(\nu)\ \frac{\tau_{\rm T} R_{\rm c }}{4\pi\rho}\
  \mbox{\rm arctg}{\sqrt{R_{\rm c}^2/\rho^2-1}},
\end{equation}
where the flux $F_{\rm A}(\nu)$ is defined by
Eq.~(\ref{eq:agn0}). The results of our calculations of the radio
excess toward a galaxy cluster (in comparison with the
background level) associated with the scattered emission of a
switched-off central galaxy using this formula are shown in
Fig.\,\ref{fig:agnbeam} as a function of $\rho$ for several
radio frequencies (black dotted lines). It can be seen that this
emission is characterized by a powerful central peak that is
narrower at high frequencies $\nu\ga1$ GHz and broader at low
ones $\nu\la500$ MHz. It follows from Eq.~(\ref{eq:agnbeam})
that the frequency dependence reflects simply the power of the
emission from the galaxy at a given frequency. The distribution
of other excess radiation components also shown in the figure,
namely the bremsstrahlung (blue dashes), the scattered radio
background (cyan solid lines), and the missing CMB due to the
scattering (long red dashes), is much flatter and smoother. All
of the components, except for the diffuse radiation from the
galaxy, are measured from the CMB decrement. The green solid
lines indicate the distribution of the combined excess
radiation. Remarkably, the central peak is retained at all
frequencies and can be used to reveal the scattered emission
from a recently switched-off central cluster galaxy (or to prove
the absence of this emission). The central emission peak is also
retained in the case where the galaxy switched off a longer time
ago. In this case, however, its amplitude decreases; moreover,
near the cluster center the intensity of the scattered emission
drops sharply. This is because the photons of the direct
radiation from the galaxy leave this region and can no longer be
involved in the scattering. Obviously, the angular radius of
this region of a reduced diffuse background is $\theta_0\sim c\,
t_{\rm so}/d_{\rm A}.$ Undoubtedly, the detected peculiarities
of the diffuse radiation morphology in the cluster are also
retained for a more realistic hot-gas density distribution than
the uniform distribution considered.

Note also that, as is obvious from Fig.\,\ref{fig:agnbeam}, the
diffuse radiation from a recently switched-off powerful central
active galaxy must maintain (and enhance) the hybrid shape of
the brightness distribution of the radio source toward the
cluster discussed above in this paper.

\subsubsection*{\uwave{Induced scattering of the emission from a
galaxy.}\/}~Above (Fig.\,\ref{fig:dev1}) we showed that the
radio background distortions due to the induced Compton
scattering manifest themselves only at very low frequencies
$\nu\la 5$ MHz. When the emission from a powerful radio galaxy
is scattered in the cluster gas, the induced scattering can
increase noticeably in importance. This is because the
intensities of the emission in the immediate vicinity of the
galaxy are high. Therefore, in particular, the question about
the possibility of gas heating near active galactic nuclei and
quasars to quasi-relativistic temperatures through the induced
Compton scattering of their radio emission was actively
discussed (Levich and Sunyaev 1971; Sazonov and Sunyaev
2001). The distortions in the spectrum of a radio source due to
the induced scattering were considered in Sunyaev (1970).

Acting in the same way as in the derivation of
Eq.~(\ref{eq:szcri}) and again using Eq.~(\ref{eq:agn0}) to
determine the flux $F_{\rm A}(\nu)$, we find the spectrum of the
scattered emission from a recently switched-off radio galaxy:
\begin{equation}\label{eq:agni}
F_{\nu}(\nu)=F_{\rm A}\, \tau_{\rm c} \frac{R_{\rm c}^2}{d_{L}^2}
\left[1-F_{\rm A}\,\frac{(1+\gamma)}{m_{\rm e}\,10^{23}}\frac{R_{\rm c}^2}{\nu^2}
\left(\frac{R_{\rm c}}{R_0}-1\right)\right]
\end{equation}
Here, $R_0$ is the radius of the region surrounding the nucleus
of the radio galaxy into which the hot intergalactic gas does
not penetrate and, therefore, in which no induced scattering
occurs. The results of our calculations using this formula are
shown in Fig.~\ref{fig:agnind} for the same cluster as that in
Fig.~\ref{fig:agn} under different assumptions about the galaxy
luminosity $L_{\rm R}=L_{41}\times 10^{41}\ \mbox{erg s}^{-1}$,
where $L_{41}=2,\ 1,\ \mbox{or}\ 0.4$ in the frequency range 10
MHz -- 100 GHz, and the radius $R_0$. It can be seen that the
scattered flux from the central radio galaxy drops sharply due to
the induced scattering already at frequencies $\sim50$ MHz and,
in a number of cases, even at $\sim100-150$ MHz, which opens the
possibility of observing the scattered radio background at lower
frequencies. Recall that the induced scattering reduces the
contribution of the scattered radio background at much lower
frequencies ($\la 5$ MHz, Fig.~\ref{fig:dev1}).

The action of the induced scattering may turn out to be even
stronger due to the scattering of the emission from a radio
galaxy by electron density fluctuations in the cluster gas. This
effect leads to a pulse broadening and rapid intensity
scintillations of radio pulsars at low frequencies and a
smearing of their images on radio maps (see, e.g., Cordes and
Lazio 1991; Bhat et al. 2004). In our case, it is important that
the scattering by electron density fluctuations lengthens the
path of the photons emitted by the radio galaxy in the cluster
gas and increases the probability of their induced scattering.

\subsubsection*{\uwave{Radiation from relativistic electrons.}\/}~Apart
from the synchrotron radiation directly associated with the
activity of cluster galaxies, the diffuse radiation (radio halo)
explained by the radiation of relativistic electrons accelerated
at shocks during cluster mergers or collisions is observed in
some clusters (Brunetti et al. 2001; van Weeren 2019). As a
rule, no radio halo is observed in relaxed clusters (Cuciti et
al. 2021). The fraction of clusters with a radio halo is
$20-50$\%, increasing with cluster mass and redshift (Cassano et
al. 2023). The halo size is comparable to the size of the X-ray
source at the location of the cluster, i.e., the width of the
bremsstrahlung profile of the hot intergalactic gas.

The lifetime of relativistic electrons can reach 10--100 Myr, so
that their radiation remains bright noticeably longer than the
diffuse (scattered in the cluster) radiation from galaxies. The
spectral slope of the synchrotron radiation from halo electrons
is almost flat at the cluster center ($\alpha\sim 0.2$) and
becomes steeper ($\alpha\sim 0.5$) on the periphery (Brunetti et
al. 2001).

Obviously, the synchrotron radiation from the halo can become an
almost unsurmountable obstacle when investigating the radio
background scattering effect in the hot cluster gas. Therefore,
it makes sense to study only well-relaxed clusters or to use
peripheral observations (the distortions due to the scattering
must have a larger scale than the halo size). Just as the
thermal bremsstrahlung flux, the nonthermal synchrotron (diffuse)
flux must decrease rapidly with redshift. Therefore, it is more
preferable to observe distant clusters.

\section*{CONCLUSIONS}
\noindent
We explored the possibility of observing the increase in the
brightness of the radio background toward galaxy clusters due to
its Compton scattering by electrons of the hot intergalactic
gas. At frequencies above $800$ MHz this scattering leads to a
decrease in the CMB brightness --- an effect predicted by
Sunyaev and Zeldovich (1970, 1972) and later confirmed in many
ground-based and space experiments. At present, this effect is
one of the main methods of observational cosmology (along with
direct X-ray and optical observations and gravitational lensing
methods). In contrast to direct observations, this effect does
not depend on the distance, allowing clusters at high redshifts
to be investigated.

In recent years, as new highly sensitive radio telescopes are
put into operation, it has been possible to also measure the
spectrum of the cosmic radio background with a high accuracy at
frequencies from tens of MHz to several GHz. It probably has a
synchrotron origin (a power-law spectrum). However, on the
whole, its nature is unclear; it is possible to associate no
more than 25\% of the measured flux with faint unresolved radio
galaxies. Nevertheless, the radio background has a high degree
of isotropy and homogeneity like the CMB. The relevant question
has arisen as to whether the radio background distortions can be
observed towards galaxy clusters due to its interaction with the
hot intergalactic gas. Observations of the distortions would
allow one to investigate both the galaxy clusters and the
properties of the radio background at high redshifts proper.

As a result of our computations, we showed the
following.

(1) The effect does exist, and at frequencies $\nu\la 800$ MHz
the excess of the radio flux above the background that is formed
due to the scattering must more than compensate for the CMB
deficit.

(2) The excess of radiation again disappears at frequencies
$\nu\la 5$ MHz due to the action of the induced Compton
scattering that carries away the photons downward along the
frequency axis.

(3)~In many cases, the intrinsic thermal (bremsstrahlung)
radiation of the hot intergalactic gas and, possibly, the
scattered synchrotron radiation associated either with the
past activity of cluster galaxies in the radio band (the
contribution of their radiation at the current level of activity
can still be taken into account and removed) or with the
radiation of relativistic electrons produced in shocks when a
given cluster merges or tidally interacts with its nearest
neighbors must hinder the direct measurement of the Compton
excess of radiation.

(4) The contribution of both bremsstrahlung and synchrotron
radiation must decrease rapidly as the cluster recedes (as its
redshift increases).

(5) In many nearby clusters the bremsstrahlung of the hot gas
completely dominates at frequencies from 10--50 MHz to 2--6 GHz;
the frequency ranges optimal for searching for and measuring the
Compton excess of the radio background flux were determined for
various parameters of clusters. Massive clusters with a hot
($kT_{\rm e}\ga 8$ keV) rarefied ($N_{\rm e}\la
10^{-2}\ \mbox{cm}^{-3}$ ) intergalactic gas at high ($z\ga0.5$)
redshifts turned out to be most promising for such observations.

(6) The peripheral observations of rich unperturbed clusters
with a density distribution that gradually falls off with radius
have certain advantages for detecting the Compton excess of the
radio background (because of the faster drop in the radio
bremsstrahlung flux).

(7) The cluster image on the radio background maps in the
frequency range 1.0--1.5 GHz must take a highly unusual
(``hybrid'') form --- a bright source at the center (associated
with the excess bremsstrahlung) surrounded by a ringlike shadow
region (a region of the CMB deficit due to the scattering); at
lower frequencies the source toward the cluster has a usual
(``positive'') form initially because of the increase in
bremsstrahlung intensity and subsequently because of the excess
of the radio background due to its scattering in the cluster
gas; at higher frequencies a shadow (``negative'') source (a
``hole'' in the background) appears at the location of the
cluster because of the CMB deficit associated with to the
scattering by electrons.

(8) The transition near a frequency of 217 GHz from the
``negative'' source on the maps of CMB fluctuations (due to the
upward escape of photons along the frequency axis upon their
scattering in the cluster gas) to the ``positive'' one (because
of the scattered photons coming to the region $\nu>217$ GHz) is
also accompanied by the appearance of a ``hybrid'' source at the
location of the cluster with a bright peak of bremsstrahlung
rising from the center of a wide Compton hole.

(9) The presence of radio galaxies in the cluster that were
active in the past can additionally complicate the observations
of the excess of the radio background due to its Compton
scattering: their synchrotron radiation scattered in the
intergalactic gas traverses a noticeably longer path (because of the
giant cluster sizes --- hundreds of parsecs) and arrives
with a noticeable delay; it is the scattered radiation that is
important, since the contribution of the current radiation from
galaxies to the total radio flux from the cluster can still be
calculated and taken into account; the scattered radiation has a
synchrotron nature and a power-law spectrum close to the
background spectrum.

(10) It is possible to measure the excess of the radio
background associated with the scattering if the luminosity of
the cluster radio galaxies and the time elapsed after their
switch-off satisfy certain constraints; we estimated them by
using the Monte Carlo method; we also showed that the diffuse
radiation of galaxies is suppressed due to the induced Compton
scattering at $\nu\la 50-100$ MHz, which opens the possibility
of observing the scattered radiation of the cosmic radio
background at these frequencies without obstruction.

(11) The morphology of the scattered radiation
from radio galaxies differs greatly both from the flat
distribution of the scattered radio background and
from the distribution of the bremsstrahlung from the
hot cluster gas; this can be used to identify the radio
excess detected toward the cluster.

In conclusion, note that the diffuse synchrotron radiation from
the relativistic electrons in the cluster gas probably formed
during the merging or the tidal effect on the cluster from its
neighbors observed as a giant radio halo can be the most serious
obstacle in the way of detecting and identifying the Compton
excess of the radio background. To detect the excess, it is
necessary to use well-relaxed clusters in which, as a rule, no
radio halo is observed and, desirably, clusters at sufficiently
high redshifts, since the intensity of the synchrotron radiation
drops rapidly with $z$.

\vspace{-0mm}

\section*{ACKNOWLEDGMENTS}
\noindent
We thank S.A. Trushkin and E.M. Churazov for their
useful remarks.

\section*{FUNDING}
\noindent
The participation of S.A. Grebenev was supported by the BASIS
Foundation for the Advancement of Theoretical Physics and
Mathematics, grant no. 22-1-1-57-1 of the Program ``Leading
Scientist (Theoretical Physics)''.

\section*{CONFLICT OF INTEREST}
\noindent
The authors of this work declare that they have no
conflicts of interest.

\vspace{5mm}

\begin{flushright}
{\sl Translated by V. Astakhov}
\end{flushright}
\end{document}